\documentclass[%
reprint,
superscriptaddress,
nofootinbib,
nolongbibliography,
 amsmath,amssymb,
 aps,
prd,
]{revtex4-1}

\usepackage{graphicx}%
\usepackage{dcolumn}%
\usepackage{bm}%
\usepackage{hyperref}%
\usepackage{xcolor}

\newcommand{\be}{\begin{equation}}
\newcommand{\ee}{\end{equation}}
\newcommand{\beqa}{\begin{eqnarray}}
\newcommand{\eeqa}{\end{eqnarray}}
\newcommand{\dd}{{\rm d}}

\DeclareMathOperator{\sinc}{sinc}

\begin{document}

\title{Quasimonochromatic LISA Sources in the Frequency Domain}%

\author{Vladimir Strokov}
\email[]{vstroko1@jhu.edu}
\affiliation{Department of Physics \& Astronomy, Johns Hopkins University}

\author{Emanuele Berti}
\email[]{berti@jhu.edu}
\affiliation{Department of Physics \& Astronomy, Johns Hopkins University}

\date{\today}%

\begin{abstract}
Among the binary sources of interest for LISA some are quasimonochromatic, in the sense that the change in the gravitational wave frequency $\Delta f\lesssim 1\;\mbox{yr}^{-1}$ during the observation time. For these sources, we revisit the stationary phase approximation (SPA) commonly used in Fisher matrix calculations in the frequency domain and show how it is modified by the Doppler shift induced by LISA's motion and by the LISA pattern functions. We compare our results with previous work in the time domain and discuss the transition from the quasimonochromatic case to the conventional SPA which applies when $\Delta f\gtrsim 1\;\mbox{yr}^{-1}$.
\end{abstract}

\maketitle

\section{Introduction\label{sec:intro}}

The Laser Interferometer Space Antenna (LISA) will be sensitive to gravitational waves (GWs) with frequencies between $\sim 10^{-4}$~Hz and $1$~Hz~\cite{2017arXiv170200786A,Seoane:2021kkk,LISA:2022yao}, thus filling the gap between the high-frequency window
covered by present and upcoming ground-based detectors (LIGO/Virgo/KAGRA, Cosmic Explorer, and the Einstein Telescope~\cite{KAGRA:2021duu,Evans:2021gyd,Maggiore:2019uih}) and the low-frequency band
accessible to Pulsar Timing Arrays~\cite{NANOGrav:2023gor,EPTA:2023fyk,Reardon:2023gzh,Xu:2023wog}. By design, LISA is a space detector in the shape of an equilateral triangle with sides of length $2.5\times 10^6$~km and a spacecraft at each vertex. The spacecraft follow Earth-like orbits such that the triangle is inclined by $60^\circ$ with respect to the ecliptic and cartwheels as the whole satellite constellation trails the Earth with a period $T=1$~yr. Time delay interferometry will be used to monitor changes induced by passing GWs. Appropriate linear combinations produce two independent GW data streams (called the ``arm~I'' and ``arm~II'' data streams below).

LISA's sensitivity to GWs is limited by noise which, in the first approximation, can be assumed to be Gaussian and fully characterized by its (one-sided) power spectral density~$S_{\rm n}(F)$ in the frequency domain (FD). Here and below we use a capital $F$ to denote Fourier frequencies, and a lowercase $f=f(t)$ for the GW frequency as a function of time. We also use a subscript $F$ to distinguish between time-domain waveforms~$h=h(t)$ and their Fourier transforms~$h_F$.

The noise power spectral density naturally leads to the definition of an inner product between the Fourier transforms of two GW signals~$h^{(1)}_F$ and~$h^{(2)}_F$ (hereafter, FD waveforms):
\be
\left(\left.h^{(1)}_F\right|h^{(2)}_F\right) = 4\,{\rm Re}\sum\limits_{\alpha={\rm I},{\rm II}}\int\limits_{0}^{+\infty}{\frac{h^{(1)}_{F,\alpha}\left(h^{(2)}_{F,\alpha}\right)^*}{S_{\rm n}(F)}\,\dd{F}}\,,\label{eq:innerFD}
\ee
where the sum is over the two LISA arms, and~$h^{(1)}_{F,\alpha}$ and~$h^{(2)}_{F,\alpha}$ are the Fourier transforms of the time-domain (TD) signals~$h^{(1)}_\alpha$ and~$h^{(2)}_\alpha$ measured by LISA. The subscript~$\alpha$ indicates that the signal recorded by each arm is different from the actual GW as a result of two effects: the Doppler variation of the GW frequency due to LISA's motion around the Sun, and the detector response encoded in the LISA pattern functions. Also, since the observation time~$T_{\rm obs}$ is limited, the signal is effectively windowed, and the infinite integration range often reduces to a narrow range of frequencies.

The inner product plays a pivotal role in assessing the detectability of a source and inferring its parameters (e.g., the distance to a binary source and the masses of the binary components). In particular, for a waveform~$h_F$ that depends on parameters~$\theta_a$, the signal-to-noise ratio (SNR) and the Fisher information matrix~$F_{ab}$ can be calculated as follows:
\be
\mbox{SNR} = (h_F|h_F)^{1/2}\,, \qquad F_{ab} = \left(\left.\frac{\partial h_F}{\partial\theta_a}\right|\frac{\partial h_F}{\partial\theta_b}\right)\,.
\ee
The SNR gives a measure of detectability, whereas the inverse of the Fisher matrix provides an estimate of the uncertainties~$\Delta\theta_a$ and correlation coefficients~$\rho_{ab}$,
\beqa
F^{-1}_{ab} &=& \rho_{ab}\Delta\theta_a\Delta\theta_b\,,
\eeqa
where $\rho_{ab}=1$ whenever $a=b$ (a parameter is always fully correlated with itself).

Although it is natural to write the inner product in the FD, the inspiral of a binary that generates the GW $h=h(t)$ happens in the TD. To relate the two, notice that the product of two waveforms (as defined above) is an $L^2$ inner product which is invariant under Fourier transform (see Appendix~\ref{app:sec:inner}). Therefore, if we define $\widetilde{h}_{\alpha}$ to be the inverse Fourier transform of the noise-weighted waveform $h_{F,\alpha}/\sqrt{S_{\rm n}/2}$, the TD counterpart of the inner product reads
\be
\left(\left.\widetilde{h}^{(1)}\right|\widetilde{h}^{(2)}\right) = \sum\limits_{\alpha={\rm I},{\rm II}}\int\limits_{0}^{T_{\rm obs}}{\widetilde{h}^{(1)}_\alpha\widetilde{h}^{(2)}_\alpha\,\dd{t}}\,. \label{eq:SNR_TD_def}
\ee
By the convolution theorem, $\widetilde{h}_{\alpha}=h_{\alpha}\star P_{\rm n}$, where $P_{\rm n}=P_{\rm n}(t)$ is the inverse Fourier transform of~$1/\sqrt{S_{\rm n}/2}$. 

This rather convoluted FD--TD relation can be simplified by invoking the stationary phase approximation (SPA). If $f=f(t)$ is the frequency drift caused by the GW inspiral, the SPA is based on the idea that the main contribution to the Fourier integrals is from integration in the vicinity of the stationary point $t$, $f(t)=F$ (for the direct Fourier transform) or, equivalently, from the stationary point $t(F)=t$ (for the inverse). For example, we can simply write $\widetilde{h}(t)=h(t)/\sqrt{S_{\rm n}(t)/2}$, where $S_{\rm n}(t)=S_{\rm n}\left[f(t)\right]$.

In this paper we revisit the use of the SPA in obtaining FD waveforms of quasimonochromatic sources (QMS). A numerous population of such sources that is of particular interest for LISA are Galactic double white dwarfs (DWD) emitting GWs at $\sim 1$~mHz (a typical median value for the population, e.g.~\cite{Korol:2021pun}). While signals from most of these binary systems will combine to produce a confusion noise~\cite{2021arXiv210801167B}, LISA will be able to resolve $\sim 10^4$ DWDs individually (see e.g. \cite{Nelemans:2001hp,Korol:2017qcx}, as well as the review article~\cite{LISA:2022yao} and references therein). A small portion of the resolved sources, the so-called verification binaries, will be known in advance from observations in the electromagnetic spectrum and will play an important role in testing LISA's performance~\cite{Stroeer:2006rx,Kupfer:2018jee,Finch:2022prg}.

Let us elaborate on the meaning of ``quasimonochromatic,'' and on the reason why a straightforward application of the SPA to such sources may not be consistent. In this paper we call a source quasimonochromatic if the number of extra cycles due to the increase in frequency during the observation time is small, say, $\lesssim 10$. More precisely, if $f(t)=f_0 + \dot{f}_0 t$ at linear order in time $t$, the GW phase in the TD reads
\beqa
\psi(t) &=& 2\pi\int\limits_0^{t}{f(t')\,\dd{t'}} = 2\pi\left(f_0 t + \frac 12 \dot{f}_0 t^2\right)\,,
\eeqa
so that the number of extra cycles accumulated over the LISA mission lifetime
\beqa
N&\equiv& \frac{\psi(T_{\rm obs})}{2\pi}-f_0 T_{\rm obs} \nonumber \\
&=& \frac 12 \dot{f}_0 T_{\rm obs}^2 \sim f_0t_c \left(\frac{T_{\rm obs}}{t_c}\right)^2 \lesssim 10\,,
\label{eq:QuasiMonoDef} \\
\Rightarrow & {} & \Delta f=\dot{f}_0T_{\rm obs}\sim f_0\frac{T_{\rm obs}}{t_c} \lesssim \frac{10}{T_{\rm obs}} \sim 1\,\mbox{yr}^{-1}\,, \label{eq:FreqDriftDef}
\eeqa
where $\Delta f$ is the total frequency drift during the observation time, and the coalescence time $t_c$ is given by~\citep{Peters:1963ux,Peters:1964zz}
\beqa
t_c &=& \frac{5\mathcal{M}}{256}\,(\pi\mathcal{M}f_0)^{-8/3} \nonumber  \\
&\approx& 2.9\;\mbox{Myr}\,\left(\frac{\mathcal{M}}{0.44M_\odot}\right)^{-5/3}\left(\frac{f_0}{2\;\mbox{mHz}}\right)^{-8/3}\sim 1\;\mbox{Myr}\,, \nonumber \\
&{}& 
\eeqa
where $\mathcal{M}$ is the chirp mass of the source and the fiducial value $0.44M_\odot$ corresponds to a $0.5M_\odot+0.5M_\odot$ binary. Here and below we use geometrical units $G=c=1$ (where $G$ is the gravitational constant, and $c$ is the speed of light).

As noted above, Eq.~(\ref{eq:QuasiMonoDef}) implies that the total change in frequency during the observation time $\Delta f\lesssim 1\,\mbox{yr}^{-1}$ is of the same order as $1/T=1\,\mbox{yr}^{-1}$, the frequency associated with LISA's motion around the Sun and encoded in the LISA Doppler phase and pattern functions. This suggests that care should be taken with the simple SPA prescription, in which one substitutes $t=t(f)$ to go to the FD. The caveat is especially evident for the LISA Doppler phase, that modifies the stationary phase condition as follows:
\be
f(t)\left[1+v(t)\right]=F \quad \Rightarrow \quad \frac{F-f}{f}=v\,,
\ee
where $v$ oscillates with the maximum amplitude $v_0\approx 30\,\mbox{km/s}/c\sim 10^{-4}$ (for a source on the ecliptic plane). Then, as long as $(F-f)/f\sim \Delta f/f_0 \lesssim 10^{-4}$, the equation has multiple solutions and, hence, multiple stationary points.

In the rest of the paper we demonstrate in detail how this multiplicity arises from the harmonics of the LISA Doppler shift factor and those of the LISA pattern functions~\cite{Cornish:2003vj} (see also the Appendix of Ref.~\cite{Cornish:2007if}), and how it affects the calculation of the SNR and of the Fisher parameter estimation errors for QMSs. In Section~\ref{sec:DWDquasimono} we summarize our assumptions and further motivate the application of our calculations to Galactic DWDs. In Section~\ref{sec:mono} we start off with the case of a perfectly monochromatic source. We first consider only the effect of the LISA Doppler modulation (which allows us to carry out the calculation fully analytically) in Section~\ref{subsec:LISADoppler}, and we include the LISA pattern functions in Section~\ref{subsec:LISApattern}. In Section~\ref{sec:quasimono} we deal with QMSs and we illustrate how the transition to the conventional SPA occurs. In Section~\ref{sec:conclusions} we summarize our results and outline directions for future work.

\section{Double white dwarfs as quasimonochromatic sources\label{sec:DWDquasimono}}

\begin{figure}[!htbp]
    \includegraphics[width=\columnwidth]{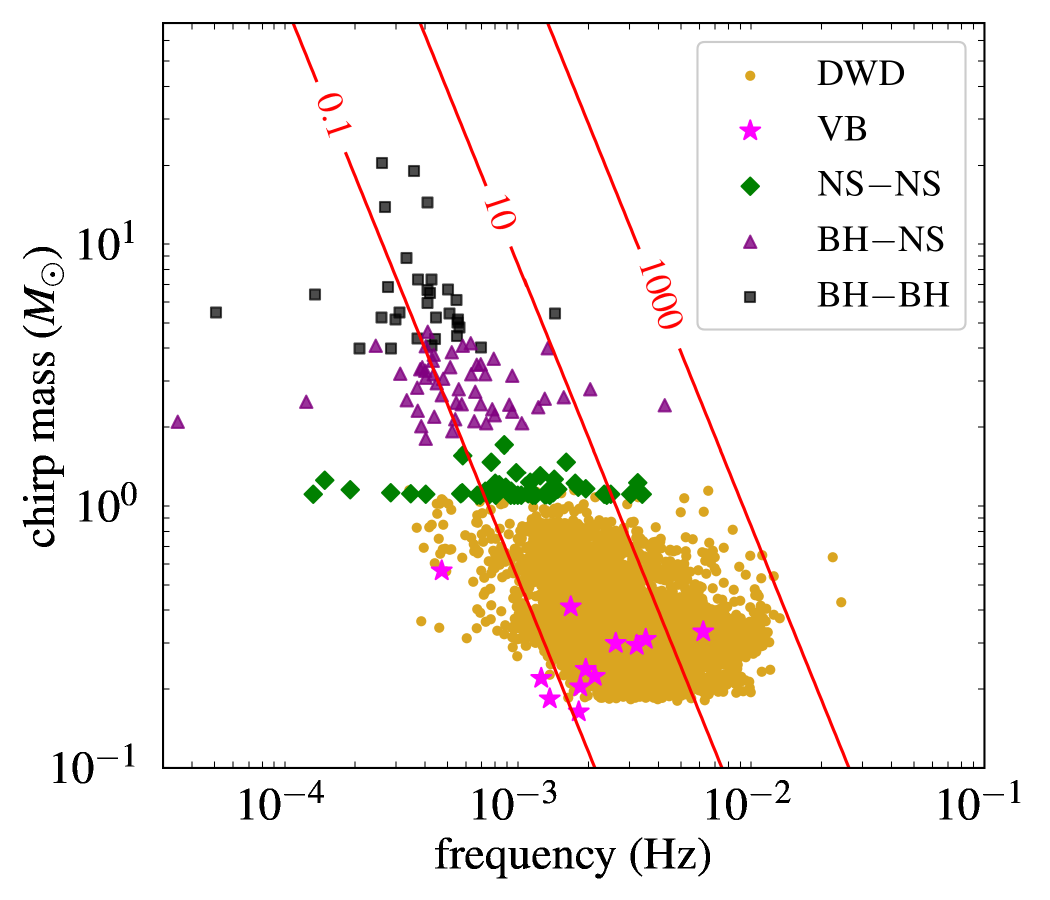}
    \includegraphics[width=\columnwidth]{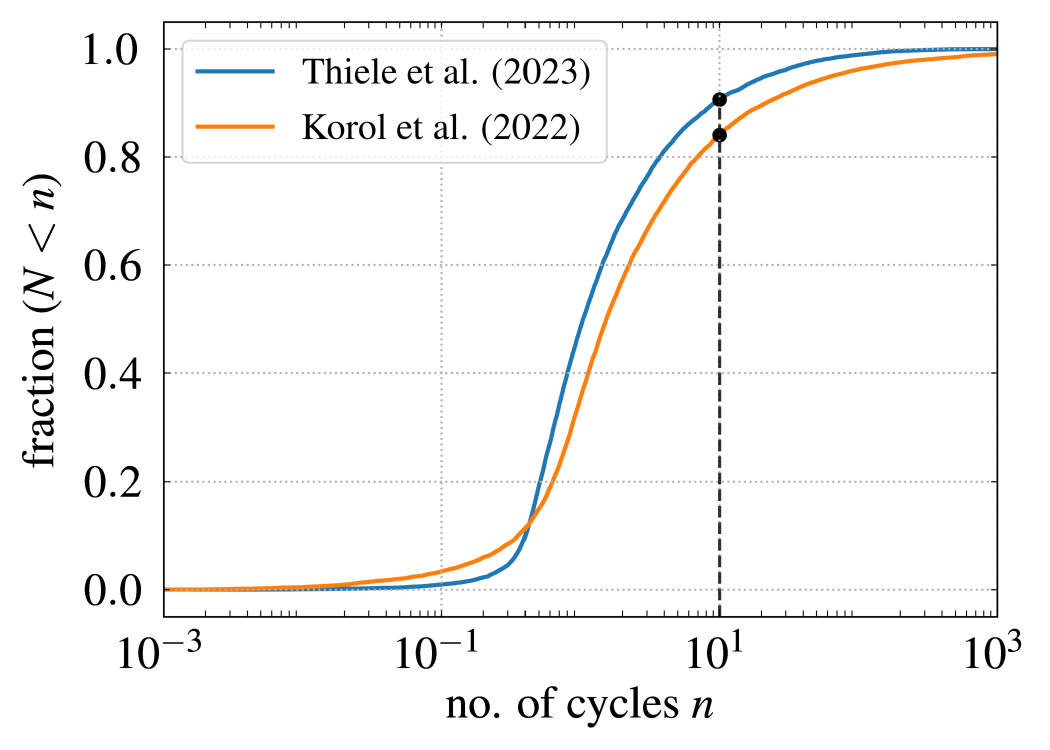}
    \caption{Top: A synthetic population of Galactic compact binaries~\cite{Thiele:2021yyb,Wagg:2021cst} in the GW frequency--chirp mass plane overlayed with three levels of the number of extra cycles accumulated due to the GW-driven frequency drift: $N=0.1$, $N=10$, and $N=1000$. The subpopulations marked in the legend are DWDs, verification binaries (VB), binary neutron stars (BNS), black hole--neutron star binaries (BH--NS), and binary black holes (BBH). Sources to the left of the level $N=10$ are QMSs in the sense used in this work: see Eqs.~(\ref{eq:QuasiMonoDef}) and~(\ref{eq:FreqDriftDef}). Bottom: Cumulative distribution function of the number of cycles for two synthetic DWD populations: model FZ from Ref.~\cite{Thiele:2021yyb} (orange line; the subpopulation depicted with golden dots in the top panel) and the population from Ref.~\cite{Korol:2021pun} (blue line). The dashed vertical line marks the fraction of DWDs that are quasimonochromatic according to our definition.
    \label{fig:DWDcycles}}%
\end{figure}

In Fig.~\ref{fig:DWDcycles} we show a synthetic population of Galactic compact binaries (model FZ from Ref.~\cite{Thiele:2021yyb}) in the frequency--chirp mass plane (top panel) as well as the corresponding cumulative distribution function for Galactic DWDs (bottom panel, orange line). In addition to the Galactic DWD population which consists of $\approx 16,000$ Galactic DWDs (including verification binaries), we also show the populations~\cite{Wagg:2021cst} of a handful of binary neutron stars (BNSs), black hole-neutron star binaries (BH-NS), and binary black holes (BBHs). The levels of constant~$N$, Eq.~(\ref{eq:QuasiMonoDef}), are shown in red, with sources to the left of the $N=10$ line being quasimonochromatic. The cumulative distribution function provides the fraction of Galactic DWDs having a number of cycles smaller than the given~$n$. The dashed line marks the point $n=10$, which corresponds to $\approx 90\%$ of the QMSs. 
Two comments are in order:
\begin{itemize}
\item[(i)] The cutoff $N=10$ we use is an order-of-magnitude estimate, but it is clear that QMSs constitute at least half of the Galactic DWDs and a significant portion of the other compact binaries. This conclusion is rather model-independent: in the bottom panel of Fig.~\ref{fig:DWDcycles}, for comparison, we show a different synthetic DWD population (from Ref.~\cite{Korol:2021pun}), which also contains a significant fraction of QMSs.
\item[(ii)] Although we calculate the number of cycles under the assumption that the binaries are detached (i.e., their inspiral is driven by GW emission), corrections to the derivative $\dot{f}_0$ induced by mass transfer and tidal interactions are comparable to its GW value (see e.g.~\cite{Breivik:2017jip,Tauris:2018kzq,Yi:2023osk}). Moreover, the non-GR corrections are typically negative, so they tend to make a GW source more monochromatic.
\end{itemize}
In any case, what follows applies to any GW sources with a small enough~$\dot{f}_0$, regardless of what process is responsible for the drift.

Note that the condition given by Eq.~(\ref{eq:QuasiMonoDef}) also implies $\Delta f/f_0\sim T_{\rm obs}/t_c\ll 1$. This justifies the use of the linear approximation for the GW frequency drift of a QMS. Indeed, to within numerical factors, the derivatives $\dot{f}_0\sim f_0/t_c$, $\ddot{f}_0\sim f_0/t_c^2\,,\ldots$\,, and the expansion of $f(t)$ is an expansion in $t/t_c\sim T_{\rm obs}/t_c$. From Eq.~(\ref{eq:QuasiMonoDef}), it then follows that $T_{\rm obs}/t_c\lesssim \sqrt{10/f_0 t_c} \sim 10^{-5} \ll 1$, and we can assume~\cite{Takahashi:2002ky,Cutler:1997ta}
\begin{itemize}
\item $S_{\rm n}\approx {\rm const}$ and $\widetilde{h}(t)\approx h(t)/\sqrt{S_{\rm n}(f_0)/2}$. That is because
\be
\frac{\Delta S_{\rm n}}{S_{\rm n}}\approx k\frac{\Delta f}{f}\ll 1\,,  \quad k \equiv \frac{\partial(\ln S_{\rm n}(f_0))}{\partial(\ln{f})}\,,
\ee
where typically the logarithmic slope $|k|\lesssim 10$ for LISA~\cite{2021arXiv210801167B}. This upper bound on $k$ can be violated at the high-frequency ``wiggles'' in the noise curve, where $k$ can increase to a few dozen. This, however, hardly affects the approximations used for QMSs, because typically $\Delta f/f\sim 10^{-5}$.
\item Similarly, the intrinsic GW amplitude of a TD waveform $A \approx {\rm const}$, since $A \propto f^{2/3}$ and the relative correction 
\be
\frac{\Delta A}{A} \sim \frac{\Delta f}{f}\ll 1\,.
\ee
\end{itemize}

In the rest of this paper we will often measure the frequency drift in $\mbox{yr}^{-1}$, the time derivative $\dot{f}$ in $\mbox{yr}^{-2}$, and, accordingly, the time in years (yr). For reference, the conversion factors are: $1\,\mbox{yr}^{-1} \approx 3\times 10^{-8}\,\mbox{Hz}$, $1\,\mbox{yr}^{-2} \approx 10^{-15}\,\mbox{s}^{-2}$, and the typical frequency $1\,\mbox{mHz}\approx 3\times 10^{4}\,\mbox{yr}^{-1}$. Also, throughout this paper we assume an observation time $T_{\rm obs}=10$~yr, which is slightly larger than the nominal mission lifetime ($T_{\rm obs}=4$~yr) but probably achievable~\cite{2017arXiv170200786A,Seoane:2021kkk,LISA:2022yao}.

\section{Monochromatic source\label{sec:mono}}

In this section we consider the case of a perfectly monochromatic source ($\dot{f}_0=0$ at all times). We start off with the simpler case in which only the LISA Doppler phase is included, and follow it up by including the LISA pattern functions. The FD decomposition into harmonics presented below is similar to the one considered in Ref.~\cite{Cornish:2003vj} (see also the Appendix of Ref.~\cite{Cornish:2007if}).

\subsection{LISA Doppler phase\label{subsec:LISADoppler}}

The TD waveform of a Doppler-modulated monochromatic source reads
\beqa
h(t) &=& A\cos{\psi(t)}\,, \label{eq:TD_waveform}\\
\psi(t) &=& 2\pi f_0 t +\psi_0 + \psi_D\,,
\eeqa
where $\psi_0$ is the initial phase offset and $\psi_D$ is the LISA Doppler contribution~\cite{Cutler:1997ta, Berti:2004bd,Robson:2018ifk}, which is given in terms of the sky location of the source $(\overline{\theta}_S,\overline{\phi}_S)$ and of LISA's angular position~$\overline{\phi}(t)$ as follows:  
\be
\psi_D = 2\pi f_0\overline{R}\cos{\left(\overline{\phi}(t)-\overline{\phi}_S\right)}\,, \quad \overline{R}\equiv R\sin{\overline{\theta}_S}
\ee 
with $R=1\,\mbox{AU}\approx 500\,\mbox{s}$ (in geometrical units). All angles refer to a coordinate system with the Solar System barycenter at the origin and the $z$ axis perpendicular to the plane of the ecliptic, so that 
\be
\overline{\phi}(t) = \overline{\phi}_0 + \frac{2\pi t}{T}\,, \quad T=1\;\mbox{yr}\,,
\ee
where $\overline{\phi}_0$ determines the position of LISA at the start of observation. Hereafter, we assume $\overline{\phi}_0=0$\,.

Let us now compute the SNR of the source and the Fisher matrix for a set of parameters $\theta_a = \left\{\ln A,\ln f_0,\psi_0,\overline{\theta}_S,\overline{\phi}_S\right\}$ in both the TD and FD. The TD calculation partially reproduces that by~\citet{Takahashi:2002ky} and, if not stated otherwise, we use the same fiducial values for the angles: $\cos{\overline{\theta}_S}=0.3$, ${\overline{\phi}_S}=5$. The TD result will serve as a consistency check for the subsequent FD calculation. 

\subsubsection{Time domain}

The SNR of the source is (see Appendix~\ref{app:sec:SNR})
\beqa
\mbox{SNR} &\approx& A\sqrt{\frac{T_{\rm obs}}{S_{\rm n}(f_0)}}\,.
\eeqa

For the Fisher matrix calculation we note that $A\,\partial h(t)/\partial A = h(t)$. Since the SNR does not depend on the subset $\theta_i=\left\{\psi_0, \overline{\theta}_S, \overline{\phi}_S\right\}$, by the property of the inner product, Eqs.~(\ref{app:eq:inner1}) and~(\ref{app:eq:inner2}), we have
\be
\left(\frac{\partial h}{\partial(\ln A)}\left|\frac{\partial h}{\partial\theta_i}\right.\right) = \left(h\left|\frac{\partial h}{\partial\theta_i}\right.\right) = 0\,.
\ee
Regarding $\ln f_0$, we modify Eq.~(\ref{app:eq:inner2}) to obtain:
\beqa
\frac{1}{\mbox{SNR}^2}\left(h\left|\frac{\partial h}{\partial(\ln{f_0})}\right.\right) &=& \frac{\partial(\ln\mbox{SNR})}{\partial(\ln{f_0})} \nonumber \\
&=& -\frac 12\frac{\partial(\ln S_{\rm n}(f_0))}{\partial(\ln{f_0})} \equiv -\frac 12 k\,. \nonumber \\
&{}& 
\eeqa
Recall that the logarithmic slope of the LISA noise curve $|k|\lesssim 10$. We find that this off-diagonal term introduces only weak correlations between $A$ and other parameters, which is why we neglect it below (roughly speaking, this term propagates the uncertainty in frequency to other parameters, which has little effect, because the frequency is measured precisely). Therefore, to within terms $\sim (f_0 T)^{-1}$, the Fisher matrix has a block structure:
\be
F_{ab} \approx \mbox{SNR}^2\times\left(
\begin{array}{cc}
1 & \boldsymbol{0} \\
{} & {} \\
\boldsymbol{0} & \displaystyle\int_0^{T_{\rm obs}}{\frac{\partial\psi}{\partial\theta_i}\frac{\partial\psi}{\partial\theta_j}\frac{\dd{t}}{T_{\rm obs}}}
\end{array}
\right)\,.
\ee
Table~\ref{tab:TD:Doppler} shows the uncertainties and correlation coefficients resulting from the inversion of the Fisher matrix. They are consistent with the values listed in Table~1 of Ref.~\cite{Takahashi:2002ky}. Note however that the case considered in that reference is somewhat different, in that it assumes a nonzero $\dot{f}_0$ and an additional pair of angles resulting from the LISA pattern functions (see also Section~\ref{subsec:LISApattern} below).

\begin{table}
\caption{Correlation matrix obtained in the TD for the case of a perfectly monochromatic source with the LISA Doppler phase included. The diagonal entries are Fisher uncertanties normalized by $\mbox{SNR}=10$, while the off-diagonal ones are the corresponding correlation coefficients $\rho_{ab}$. Also shown is the localization error $\Delta\Omega_{\rm S}=2\pi\sin{\overline{\theta}_S}\Delta\overline{\theta}_S\Delta\overline{\phi}_S\sqrt{1-\rho^2_{\overline{\theta}_S\overline{\phi}_S}}$. The fiducial parameters used in the calculation are: $f=2$~mHz, $T_{\rm obs}=10$~yr, $\cos{\overline{\theta}_S}=0.3$, ${\overline{\phi}_S}=5$\,. The values in parentheses are from a similar case studied in Ref.~\cite{Takahashi:2002ky} (see Table~1 and Figure~2 in that reference).\label{tab:TD:Doppler}}
\begin{center}
\begin{tabular}{c|ccccc}\hline\hline\\
{}  & $\Delta A/A$  & $\Delta f_0 T_{\rm obs}$  & $\Delta\psi_0$  & $\Delta\overline{\theta}_S$ & $\Delta\overline{\phi}_S$ \\
\\ \hline \\
$\Delta A/A$  & 0.1 (0.2) & 0 & 0 & 0 & 0 \\
$\Delta f_0 T_{\rm obs}$ & {} & $0.056$ ($0.055$) & $-0.87$ & 0.075 & $-0.022$ \\
$\Delta\psi_0$  & {} & {} & 0.20 & $-0.065$ & 0.019 \\
$\Delta\overline{\theta}_S$ & {} & {} & {} & 0.075 & $-0.0017$ \\
$\Delta\overline{\phi}_S$ & {} & {} & {} & {} & 0.024 \\
\hline\\
{} & {} & \multicolumn{4}{l}{Localization: $\Delta\Omega_{\rm S}=0.01\;[\mbox{sr}]\approx 35\;[\mbox{deg}^2]$}   \\
{} & {} & \multicolumn{4}{l}{\hphantom{Localization:}$\left(\Delta\Omega_{\rm S}\approx 0.01\;[\mbox{sr}]\right)$}   \\
\\
\hline
\end{tabular}
\end{center}
\end{table}

\subsubsection{Frequency domain\label{subsubsec:FD}}

Since the FD inner product, Eq.~(\ref{eq:innerFD}), contains only positive frequencies $F$, in the Fourier transform
\beqa
h_F &=& \frac 12 Ae^{i\overline{\psi}_0}\int\limits_0^{T_{\rm obs}}{\dd{t}\,e^{-2\pi i(F-f_0)t + i\psi_D(t)}} \nonumber \\
&+& \frac 12 Ae^{-i\overline{\psi}_0}\int\limits_0^{T_{\rm obs}}{\dd{t}\,e^{-2\pi i(F+f_0)t - i\psi_D(t)}}\,,
\eeqa
we can retain only the first integral. The second integral almost vanishes outside of a vicinity of $F=-f_0<0$ and is of the order of $\mathcal{O}[(f_0 T)^{-1}]$ at $F=f_0$ (as it is evident from the result of the calculation below).

Using the expansion of the Doppler modulation in terms of Bessel functions
\beqa
e^{2\pi i f_0\overline{R}\cos{\left(\overline{\phi}-\overline{\phi}_S\right)}} &\equiv& \sum\limits_m{a_m e^{im\overline{\phi}}} \nonumber \\
&=& \sum\limits_{m=-\infty}^{+\infty}{i^m J_m\left(2\pi f_0\overline{R}\right)e^{im(\overline{\phi}-\overline{\phi}_S)}}\,, \nonumber \\
&{}& \label{eq:DopplerFourier}
\eeqa
it is straightforward to obtain:
\beqa
&{}& h_F =\frac 12 AT_{\rm obs} e^{i\overline{\psi}_0}\sum\limits_{m}{h_F^{(m)}}\,, \nonumber \\
h_F^{(m)} &\equiv& a_m u_m(\nu)\,, \qquad \nu\equiv (F-f_0)T\,, \label{eq:mono_sum}
\eeqa
with
\beqa
&{}&u_m(\nu)= e^{-i\pi K\nu}\frac{\sin\left(\pi K(m-\nu)\right)}{\pi K(m-\nu)}\,, \label{eq:uFunctions}\\
&{}& a_m = i^m J_m(2\pi f_0\overline{R})e^{-im\overline{\phi}_S}\,, \label{eq:aCoefs} 
\eeqa
where we have used that $K\equiv T_{\rm obs}/T$ is an even integer.
Similar Bessel function expansions are also of use in the treatment of eccentric GW sources (see e.g.~\cite{Pierro:2000ej}).

Before we use this expansion, it is useful to make two observations. First, the tail of this sum at the negative $F = -f_0$ ($|\nu|=2f_0 T$) is indeed $\sim (f_0 T)^{-1}$ (the same argument applies as in Appendix~\ref{app:sec:SNR}). Second, the above expansion can be be viewed as a result of a triple convolution: namely, the function prior to the Fourier transform is the product of a signal $g=e^{2\pi i f_0t}$, a rectangular window $W$ that vanishes outside of $[0,T_{\rm obs}]$, and the periodic function $e^{i\psi_{D}(t)}$. The expansion~(\ref{eq:mono_sum}) then arises as follows:
\be
(gW)_F = g_F\star W_F = T_{\rm obs }e^{-i\pi K\nu}\frac{\sin\left(\pi K\nu\right)}{\pi K\nu}\,, \label{eq:windowedFDsignal}
\ee
\beqa
\left(e^{i\psi_{D}(t)}gW\right)_F  &=& \left(e^{i\psi_{D}(t)}\right)_F\star (gW)_F \nonumber \\
&=& \sum\limits_m{a_m\times (gW)_{F-m/T}} \nonumber \\
&=& T_{\rm obs}\sum\limits_m{a_m e^{-i\pi K\nu}\frac{\sin\left(\pi K(\nu-m)\right)}{\pi K(\nu-m)}}\,. \nonumber \\
&{}& \label{eq:tripleConvolution}
\eeqa

Now, one useful way to interpret the series just obtained is to notice that it is an expansion with respect to the set of orthogonal functions~$u_m(\nu)$:
\be
K\int\limits_{-\infty}^{+\infty}{\dd{\nu}\,u_m(\nu)u_{m^\prime}^*(\nu)} = \delta_{mm'}\,,
\ee
which means that the different terms contribute to the SNR independently:
\beqa
\mbox{SNR}^2&=&4\int_{-\infty}^{+\infty}{\frac{|h_F|^2}{S_{\rm n}(f_0)}\,\dd{F}} = \frac{4}{T}\int_{-\infty}^{+\infty}{\frac{|h_F|^2}{S_{\rm n}(f_0)}\,\dd\nu} \nonumber \\
&=& \frac{\left(AT_{\rm obs}\right)^2}{S_{\rm n}(f_0)}\frac{1}{KT}\sum\limits_m{J_m^2(x)} = \frac{A^2 T_{\rm obs}}{S_{\rm n}(f_0)}\,, \label{eq:mono_harmonics}
\eeqa
where we have denoted $x = 2\pi f_0\overline{R}$ and used a property of the Bessel functions. By coincidence, an estimate of the integrals for separate values of~$m$ yields the same result. Namely, the width of each harmonic is $\Delta\nu\sim 1/K$ (the first zero of the sinc function), or $\Delta F \sim 1/(KT)$, and therefore we have $|h_F^{(m)}|\Delta F\sim J_m^2(x)/(KT)$, which coincides with the result after exact integration. (Note also that $\Delta F\sim 1/T_{\rm obs}$ is roughly the width of a single frequency bin, which in practice makes the individual peaks delta-like.)

\begin{figure}[!htbp]
    \includegraphics[width=\columnwidth]{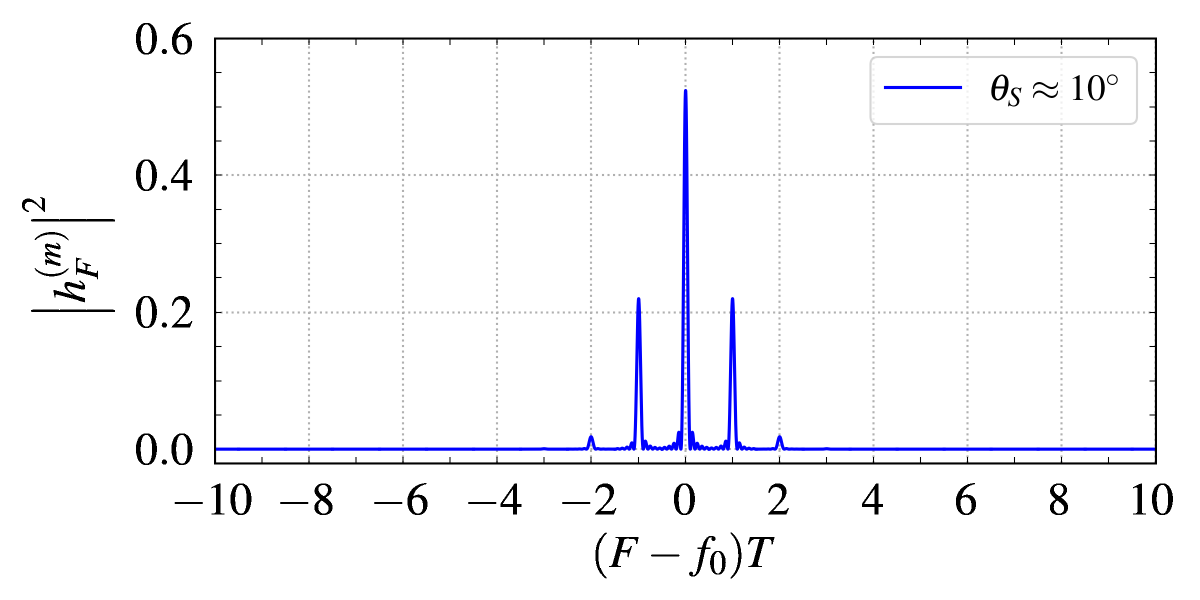}
    \includegraphics[width=\columnwidth]{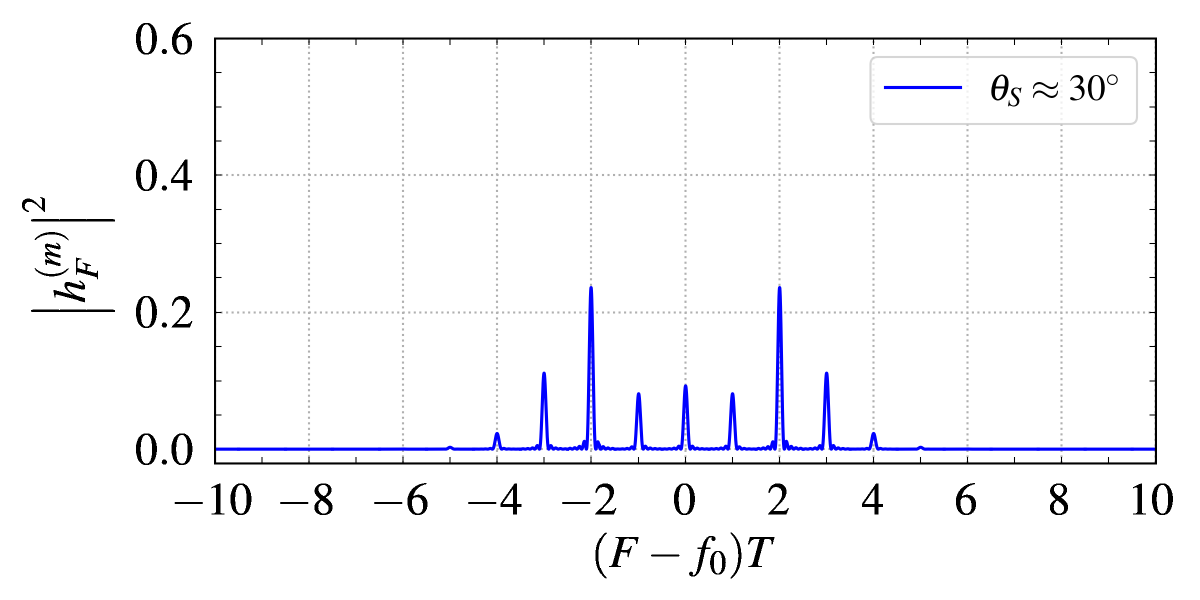}
    \includegraphics[width=\columnwidth]{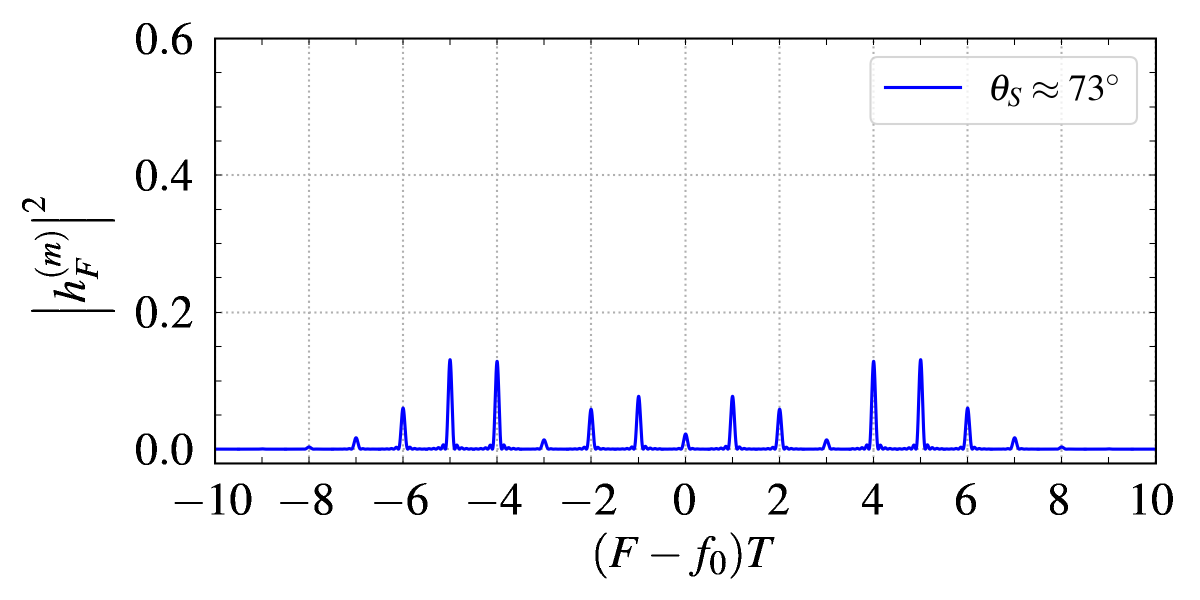}
    \includegraphics[width=\columnwidth]{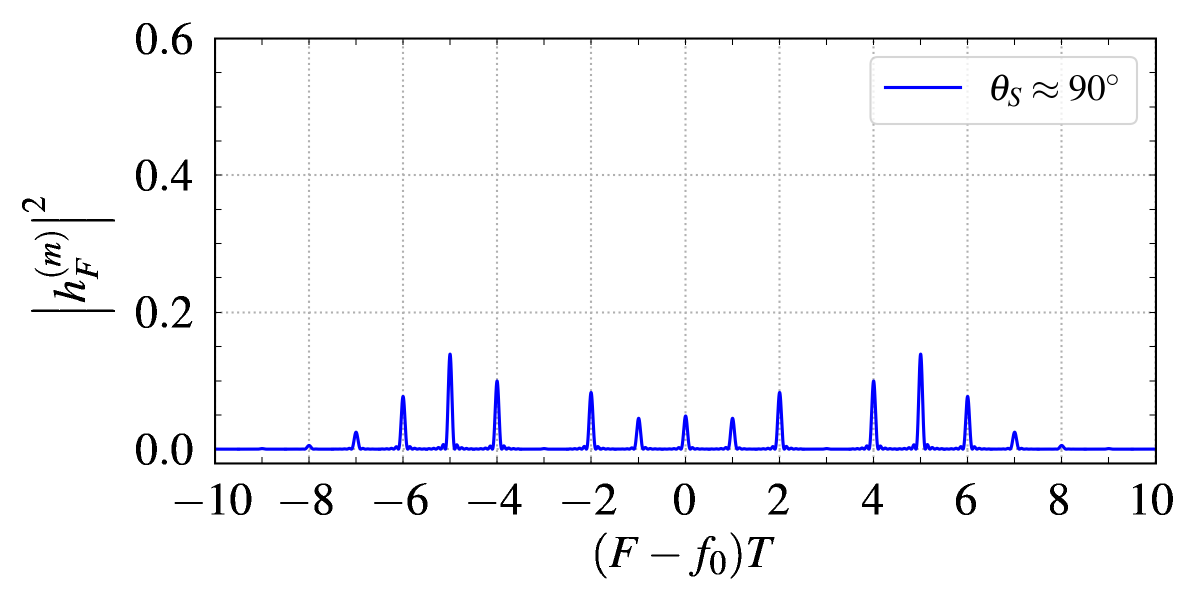}
    \caption{Power spectrum of the FD waveform of a monochromatic source with the LISA Doppler phase contribution, Eqs.~(\ref{eq:mono_sum})--(\ref{eq:aCoefs}), for four values of the parameter $x=2\pi f_0 R\sin{\overline{\theta}_S}$. From top to bottom: $x\approx 1$ ($\overline{\theta}_S=10^\circ$), $x\approx 3$ ($\overline{\theta}_S=30^\circ$), $x\approx 6$ ($\overline{\theta}_S=\arccos{0.3}$), and $x\approx 2\pi$ ($\overline{\theta}_S=90^\circ$). The fiducial value $f_0=2$~mHz is assumed, and $R=1\;\mbox{AU}/c\approx 500$~s. 
    \label{fig:mono_harmonics}}%
\end{figure}

What we can learn from Eq.~(\ref{eq:mono_harmonics}) is that, \textit{unless a source is close to a celestial pole, the contributions to the SNR budget from harmonics beyond the ``natural'' ones $(F-f_0)T=0,\pm 1$ cannot be neglected}. Moreover, as we will now demonstrate, the bandwidth of the signal exceeds the typically assumed value of $\Delta F\approx 4/T$ (see Ref.~\cite{Cornish:2003vj}) already at moderate angles~$\overline{\theta}_S$. Figure~\ref{fig:mono_harmonics} illustrates how the total SNR$^2$ is distributed over the harmonics (up to multiplication by the width of a peak). The angle~$\overline{\theta}_S$, and hence the parameter $x=2\pi f_0R\sin{\overline{\theta}_S}$ (with $f_0=2\,\mbox{mHz}$), increase from top to bottom. Already at $\overline{\theta}_S = 10^\circ$ ($x\approx 1$) the harmonics $m=\pm 2$ start emerging, and they become dominant at $\overline{\theta}_S = 30^\circ$ ($x\approx 3$), with a noticeable contribution from the harmonics $m=\pm 3$. The further increase to the fiducial value $\overline{\theta}_S=\arccos{0.3}\approx 73^\circ$ ($x\approx 6$) pushes the ``comb teeth'' as far as $m=\pm 6$, while the main harmonic $m=0$ almost vanishes. For comparison we also show the case $\overline{\theta}_S=90^\circ$ (maximum~$x$), which is qualitatively similar to the previous one.

The fact that the higher harmonics have a non-negligible contribution to the total power manifests itself in the Fisher matrix as well. For example we have
\beqa
\frac{\partial h_F}{\partial\overline{\phi}_S} &=&\frac 12 AT_{\rm obs} e^{i\overline{\psi}_0}\sum\limits_m{(-im)a_m u_m(\nu)}\,, \\
F_{\overline{\phi}_S\overline{\phi}_S} &=& \mbox{SNR}^2\sum\limits_m{m^2J_m^2(x)}=\mbox{SNR}^2\times\frac{x^2}{2}\,,
\eeqa
where the factor~$m^2$ in the sum suppresses lower harmonics and amplifies higher harmonics. For the case of $\overline{\theta}_S\approx 73^\circ$, in particular, this leads to the power being spread around $m=\pm 5$ for this specific element of the matrix. 

\begin{figure}[t]
    \includegraphics[width=\columnwidth]{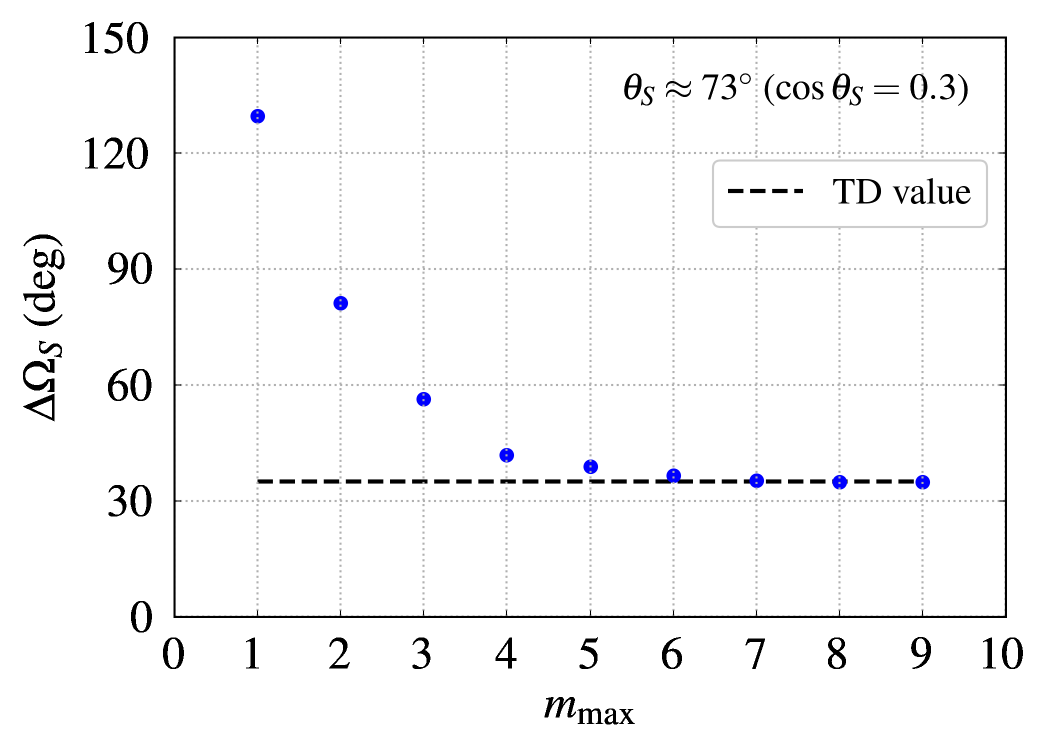}
    \caption{Localization error of the source in the FD as a function of the number of harmonics included in Eqs.~(\ref{eq:mono_sum}) and~(\ref{eq:mono_harmonics}) for the Fisher matrix calculation. Each point corresponds to integration over the peaks with $m=-m_{\rm max},\ldots,m_{\rm max}$ in the panel corresponding to $\overline{\theta}_S\approx 73^\circ$ of Fig.~\ref{fig:mono_harmonics}. The dashed horizontal line shows the TD value from Table~\ref{tab:TD:Doppler} (see the table caption for a list of the fiducial parameters). 
    \label{fig:mono_localization}}%
\end{figure}

In order to demonstrate the effect of higher harmonics, let us compute the Fisher uncertainties using only terms up to $|m|= m_{\rm max}$ in Eqs.~(\ref{eq:mono_sum}) and~(\ref{eq:mono_harmonics}). Since the uncertainty in $f_0$ is small (see Table~\ref{tab:TD:Doppler}), we can as well exclude it from the list of parameters to simplify the calculation. We have already verified in the TD that, as expected, doing so hardly affects the angular uncertainties and, thus, the localization error $\Delta\Omega_S$.

In Fig.~\ref{fig:mono_localization} we show the localization error for $\overline{\theta}_S=\arccos{0.3}$ and~$\mbox{SNR}=10$ as a function of the highest included harmonic,~$m_{\rm max}$. It is clear that the more harmonics are included, the closer the uncertainty is to the TD value (marked by the dashed horizontal line). If only the lowest harmonics $m=0, \pm 1$ are taken into account, the localization estimate is extremely off.

\subsection{LISA pattern functions\label{subsec:LISApattern}}

The LISA pattern functions $\mathcal{F}_{+,\times}$ encode the detector response to the plus and cross polarizations of a GW:
\beqa
h_\alpha(t) &=& A\cos{\psi(t)}(1+\cos^2{\iota})\mathcal{F}^{+}_\alpha  \nonumber \\
&+& 2A\sin{\psi(t)}\cos{\iota}\,\mathcal{F}_\alpha^{\times}\,, \\
\mathcal{F}_\alpha^{+,\times}&=&\mathcal{F}_\alpha^{+,\times}(t,\overline{\theta}_S,\overline{\phi}_S, \varphi)\,, \quad \alpha=\mbox{I},\mbox{II}\,, \label{eq:pattern1}
\eeqa
where $\iota$ is the orbital inclination, and $\varphi$ is the polarization angle~\cite{Isi:2022mbx}. To simplify notation, we omit the subscript~$\alpha$ labeling the LISA arms, and we focus on arm~I for the plots in this section. Results for arm~II are qualitatively the same.

The introduction of the LISA pattern function results in an extra factor in the computation of the Fourier transform (see Section~\ref{subsubsec:FD}):
\be
\mathcal{F} = (1+\cos^2{\iota})\mathcal{F}^{+} - 2i\cos{\iota}\,\mathcal{F}^{\times}\,. \label{eq:patternTD}
\ee 
Being periodic in $\overline{\phi}=2\pi t/T$ with a period of~$2\pi$, this function can be decomposed into a Fourier series: 
\beqa
\mathcal{F} &=& \sum\limits_m{b_m e^{im\overline{\phi}}}\,, \qquad \mathcal{F}^{+,\times} = \sum\limits_m{b_m^{+,\times} e^{im\overline{\phi}}}\,,
\eeqa
\be
b_m = b_m^{+}(1+\cos^2{\iota}) - 2i b_m^\times \cos\iota\,. \label{eq:patternFourier}
\ee
Then, the Fourier series for the product of the Doppler and pattern functions factors is
\beqa
\mathcal{F} e^{i\psi_D(\overline{\phi})} &=& \sum\limits_m{\widetilde{a}_m}e^{im\overline{\phi}}\,, \\
\widetilde{a}_m &=& \sum\limits_n{a_n b_{m-n}} = \sum\limits_n{a_{m-n} b_n}\,, \label{eq:discreteConvolution}
\eeqa
which is the discrete version of a convolution. 

\begin{figure}[!htbp]
    \includegraphics[width=\columnwidth]{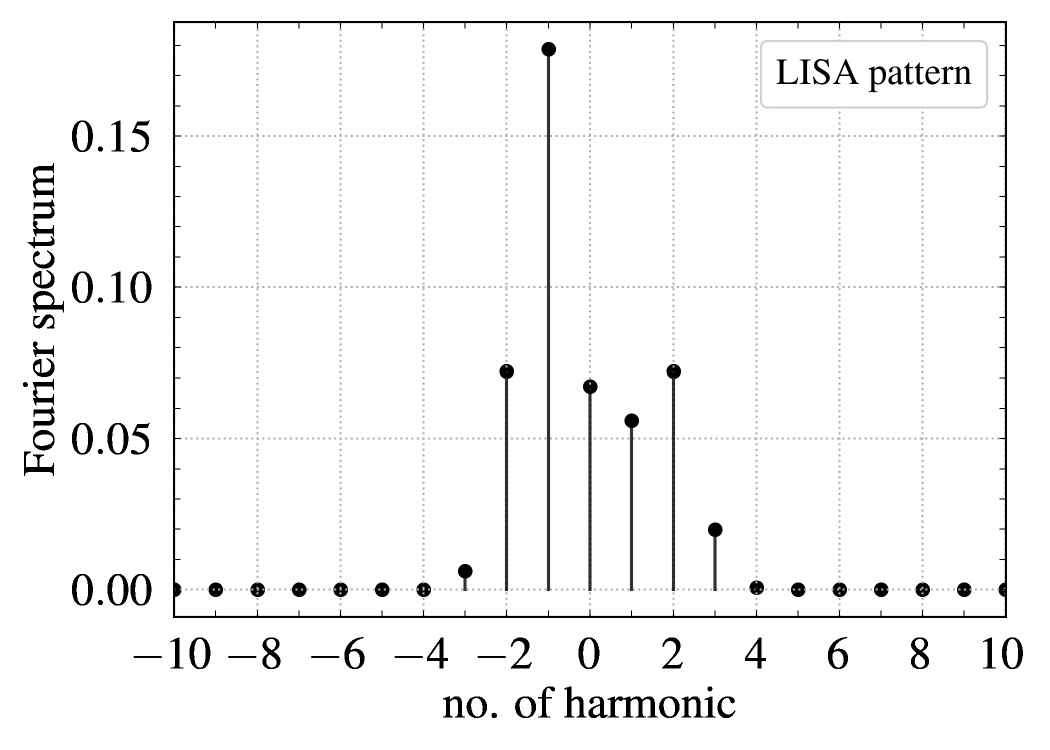}
    \includegraphics[width=\columnwidth]{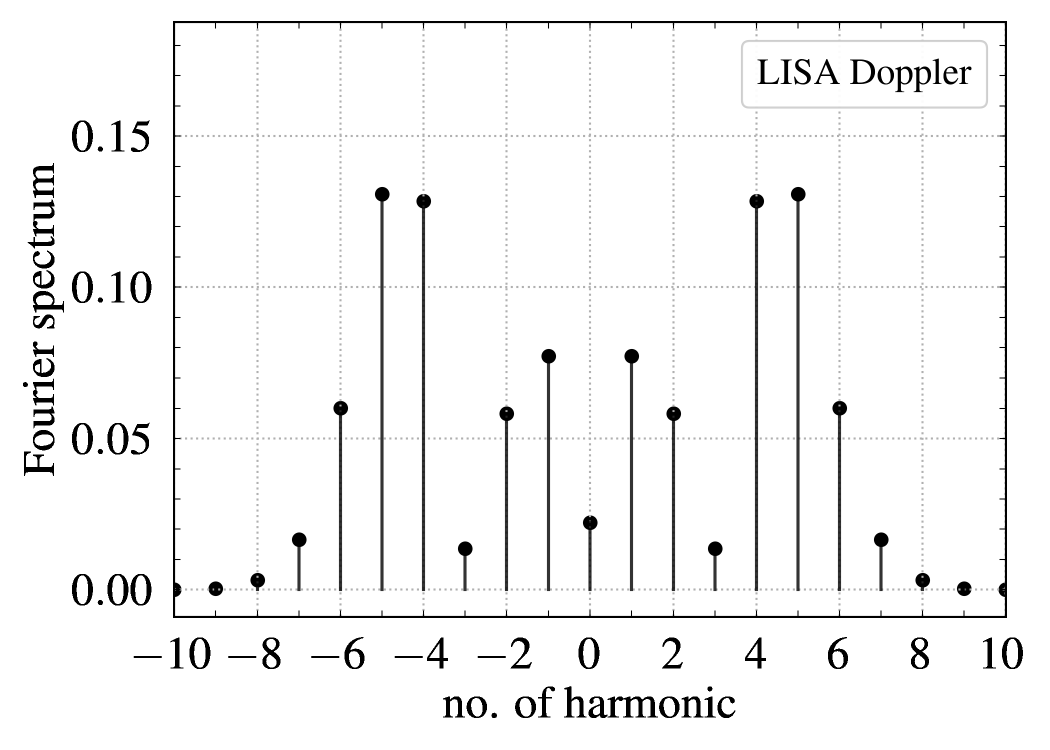}
    \includegraphics[width=\columnwidth]{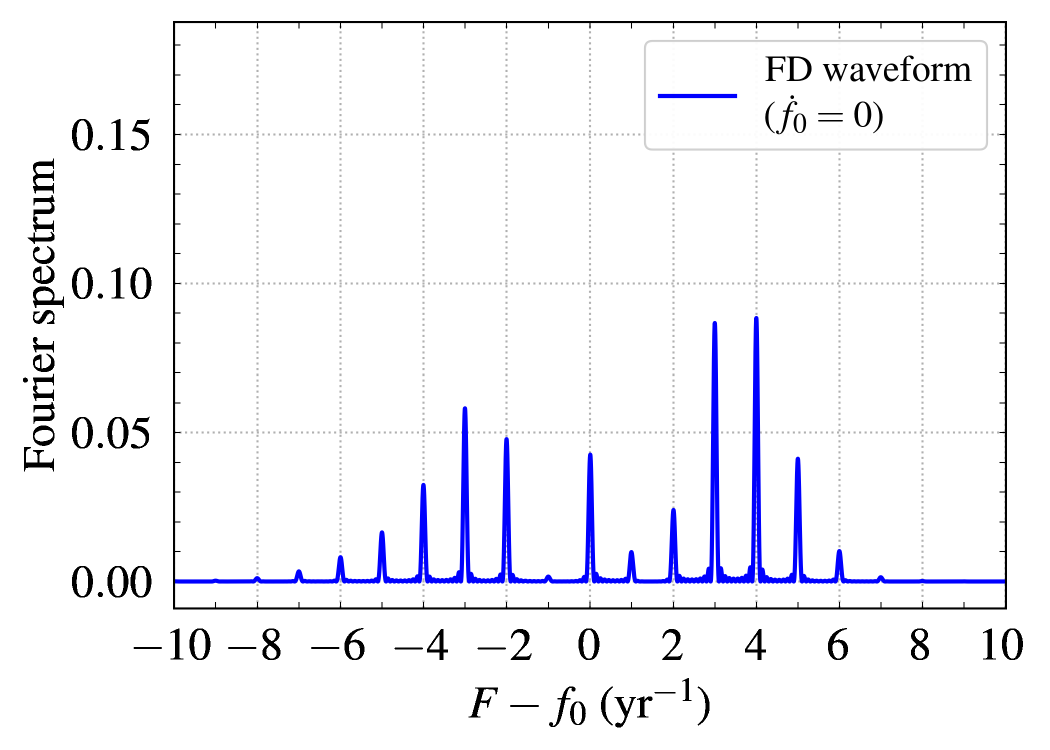}
    \caption{Fourier spectra of the LISA pattern functions (top panel, Eqs.~(\ref{eq:pattern1})--(\ref{eq:patternFourier})), the LISA Doppler factor (middle panel, Eq.~(\ref{eq:DopplerFourier})), and the FD waveform which results from the convolution of the two (bottom panel: cf. Eqs.~(\ref{eq:discreteConvolution}) and~(\ref{eq:mono_convolved_waveform})). The fiducial values for the angles are chosen to match those of Ref.~\cite{Takahashi:2002ky}: $\overline{\theta}_S=\arccos{0.3}$, $\overline{\phi}_S=5$, $\overline{\theta}_L=\arccos{(-0.2)}$, and $\overline{\phi}_L=4$ (or, $\iota\approx 60^\circ$ and $\varphi\approx 150^\circ$).  
    \label{fig:mono_LISA}}%
\end{figure}

Therefore, to include the LISA pattern functions in Eqs.~(\ref{eq:mono_sum})--(\ref{eq:aCoefs}), we simply substitute~$a_m\to \widetilde{a}_m$:
\be
\widetilde{h}_F = \frac 12 AT_{\rm obs}e^{i\psi_0}\sum\limits_m{\widetilde{a}_m u_m(\nu)}\,.
\label{eq:mono_convolved_waveform}
\ee
Unlike for the LISA Doppler factor, the Fourier coefficients~$b_m$ for the LISA pattern functions do not appear to have a closed analytic form. However, their numerical computation is straightforward, and the generic behavior of their relative amplitudes on the source angles can be easily studied (see Appendix~\ref{app:sec:patternNumerical}). 

Figure~\ref{fig:mono_LISA} shows the individual discrete Fourier spectra of the LISA pattern functions (top panel) and the LISA Doppler factor (middle panel). Their convolution~$\widetilde{a}_m$ is, in turn, convolved with the windowed FD signal, Eq.~(\ref{eq:windowedFDsignal}), to yield the full FD waveform of a monochromatic source (bottom panel). The asymmetry in the $\pm m$ harmonics of the LISA pattern functions (see also Eq.~(\ref{eq:patternFourier})) is due to the fact that they are nontrivial superpositions of the symmetric plus and cross harmonics ($|b^{+}_m|=|b^{+}_{-m}|$, $|b^{\times}_m|=|b^{\times}_{-m}|$). It is evident from this figure that the introduction of the LISA response does not change the main conclusion of the previous section: \textit{multiple harmonics induced by LISA's motion contribute to the power in the FD and, thus, must be taken into account for a consistent SNR and Fisher matrix calculation}.

\section{Quasimonochromatic source\label{sec:quasimono}}

Qualitatively, the introduction of a slight frequency drift only leads to the widening of the FD peaks compared to the monochromatic case of Fig.~\ref{fig:mono_LISA} (bottom panel). At least, this is what we can expect if the widening does not exceed the separation between the peaks, $\Delta f\lesssim 1/T=1\,\mbox{yr}^{-1}$, which is guaranteed by our definition of QMSs, Eqs.~(\ref{eq:QuasiMonoDef}) and~(\ref{eq:FreqDriftDef}). Equivalently, this condition can be expressed as
\beqa
\Delta f&=&\dot{f}_0 T_{\rm obs} \lesssim \frac{1}{T} \quad \Rightarrow \quad \dot{f}_0 \lesssim \frac{1}{KT^2}\sim 0.1\;\mbox{yr}^{-2}\,. \nonumber \\
&{}&
\eeqa
Of course, this upper limit on $\Delta f$ is approximate, and somewhat sensitive to numerical factors and to the observation time. This qualitative picture holds better for small frequency drifts.

\begin{figure}[t]
    \includegraphics[width=\columnwidth]{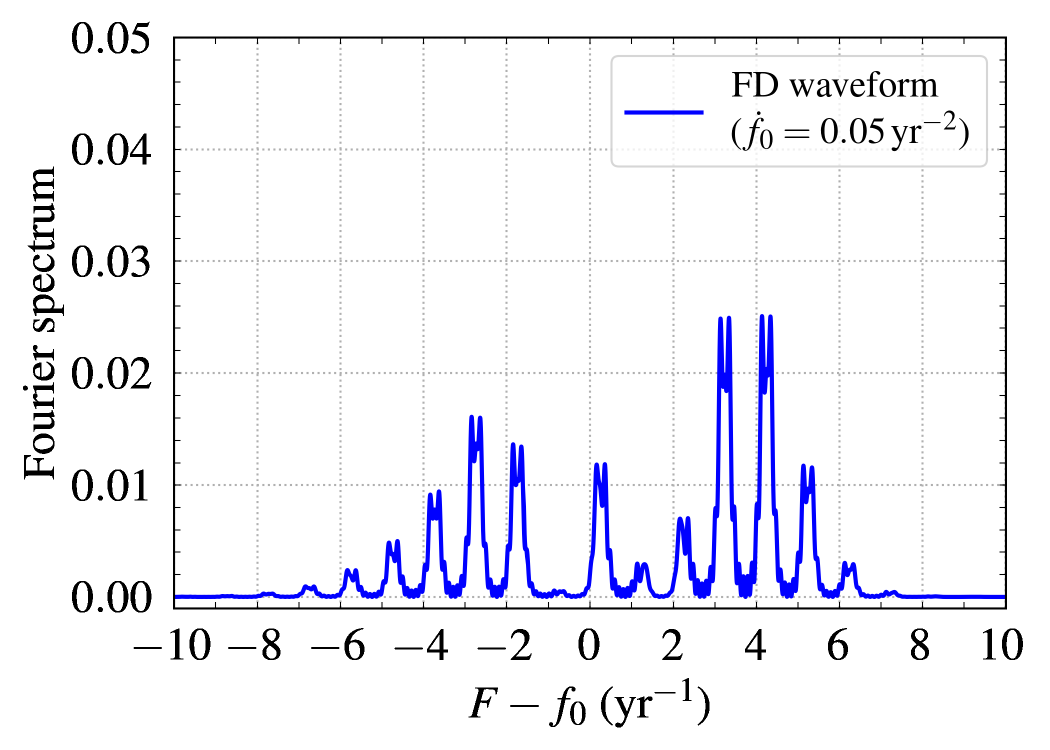}
    \caption{Fourier spectrum of a QMS FD waveform with the LISA Doppler and pattern modulation taken into account. The GW frequency drift is assumed to be linear in time with $\dot{f}_0=0.05\;\mbox{yr}^{-2}$ (e.g. a $0.6M_\odot+ 0.6M_\odot$ DWD emitting at $f_0\approx 2.5$~mHz). This is a weakly nonmonochromatic counterpart of the FD waveform shown in Fig.~\ref{fig:mono_LISA} (bottom panel).
    \label{fig:quasi_LISA}}%
\end{figure}

Quantitatively, the FD waveform obtained as a convolution of the Fourier transforms of all relevant factors (see Eqs.~(\ref{eq:tripleConvolution}), (\ref{eq:discreteConvolution}), and (\ref{eq:mono_convolved_waveform})) is readily generalized to a version of the fast-slow decomposition~\cite{Cornish:2007if}:
\be
\widetilde{h}_F = \frac 12 AT_{\rm obs}e^{i\psi_0}\sum\limits_m{\widetilde{a}_m v_{\nu-m}}\,,
\label{eq:quasi_convolved_waveform}
\ee
where $v_\nu$ is the Fourier transform of a windowed GW signal (modulo the initial phase) with the linear frequency drift:
\be
v_{\nu} = \frac{Z\left(\xi+y\right)+Z\left(\xi-y\right)}{2\xi}\,e^{-\frac 12i\pi(y+\xi)^2}\,, \label{eq:fdotFresnel}
\ee
\be
y \equiv \frac{\nu K}{\sqrt{N}}-\sqrt{N}\,, \qquad \xi \equiv \sqrt{N} = \frac 12 K\sqrt{2\dot{f}_0T^2}\,,  
\ee
and $Z(y)\equiv\int\limits_0^y{e^{\frac{i\pi z^2}{2}}\dd{z}}\equiv C(y) + iS(y)$\,, with~$C$ and~$S$ being the Fresnel integrals. Note that, since $T=1\,\mbox{yr}$, the dimensionless combination $\dot{f}_0 T^2$ is, in fact, the time derivative in yr$^{-2}$.

In Fig.~\ref{fig:quasi_LISA} we show the amplitude squared of the FD waveform of a QMS with $\dot{f}_0=0.05<K^{-1}\,\mbox{yr}^{-2}$. As expected, the main difference between this waveform and its perfectly monochromatic counterpart in Fig.~\ref{fig:mono_LISA} are the wider peaks.

\begin{figure}[t]
    \includegraphics[width=0.95\columnwidth]{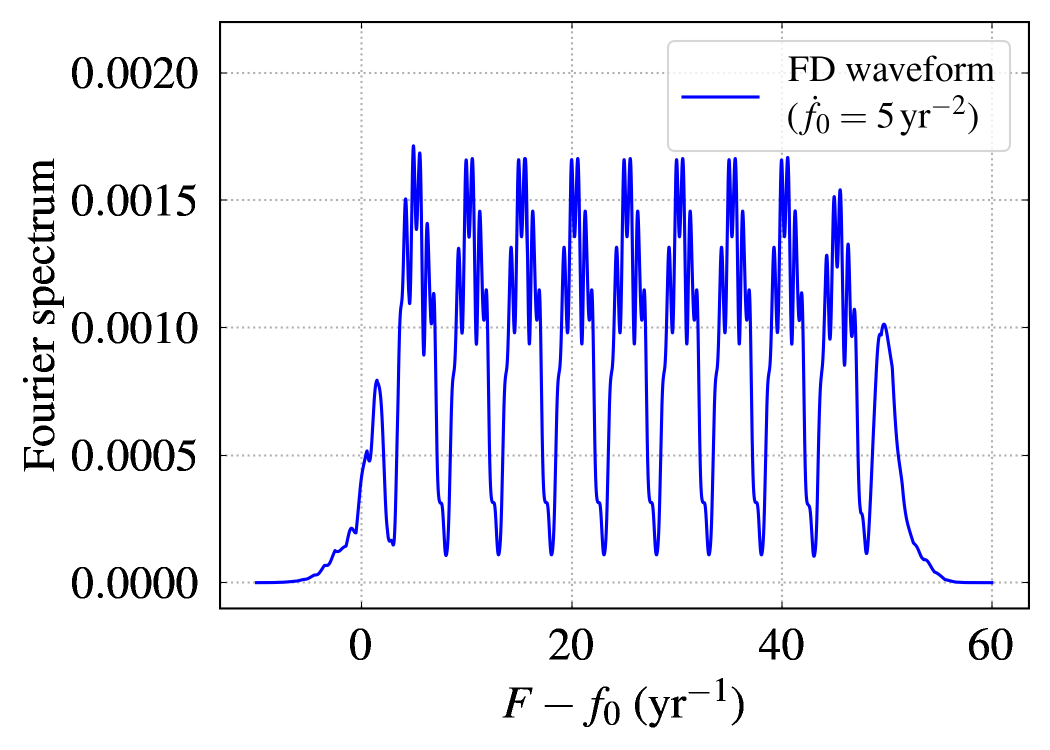}
    \includegraphics[width=0.95\columnwidth]{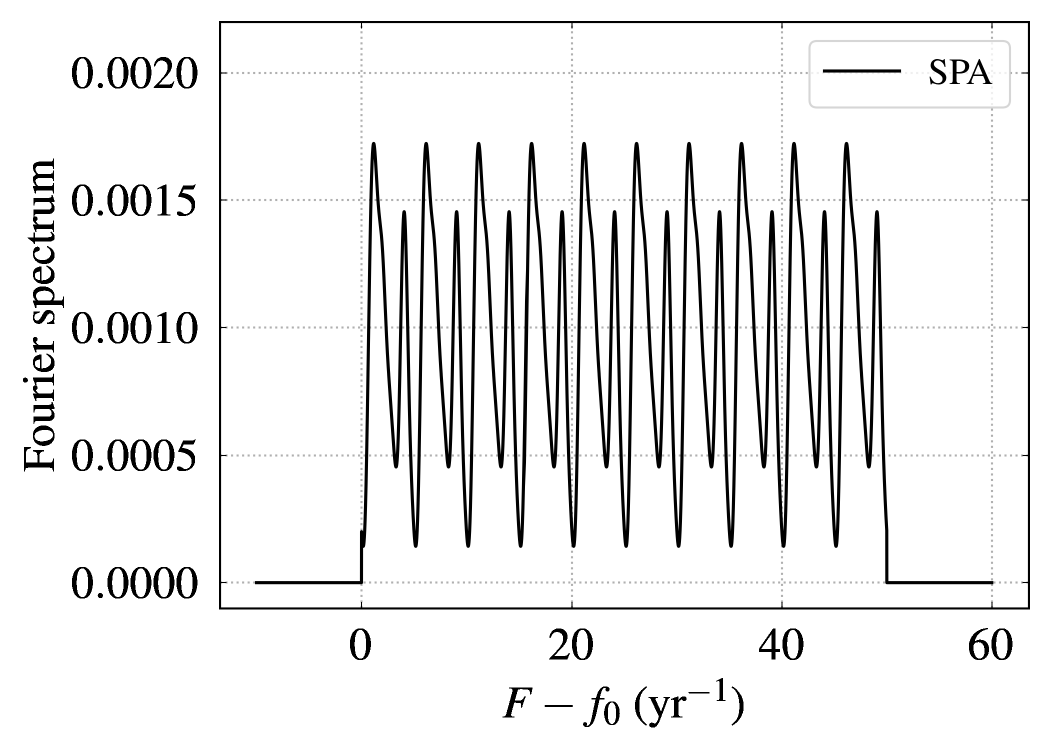}
    \caption{Comparison of the FD waveform obtained by convolution over individual harmonics of the LISA detector response (top) and its SPA counterpart (bottom). It is assumed that $\dot{f}_0=5\;\mbox{yr}^{-2}$ (e.g. a $0.6M_\odot+ 0.6M_\odot$ double WD emitting at $f_0\approx 9$~mHz), which results in a total frequency drift well above the separation between harmonics ($1\,\mbox{yr}^{-1}$). 
    \label{fig:spa_LISA}}%
\end{figure}

We now want to explore how summation over the individual harmonics of the LISA Doppler and pattern factors in Eq.~(\ref{eq:quasi_convolved_waveform}) translates into the conventional SPA when the drift $\Delta f\gg 1/T$. In that limit the FD waveform $v_\nu$ is so wide that it spans multiple harmonics. When evaluated at the shifted frequencies $(\nu-m)$, it will vary only slightly, with the main contribution coming from the variation of its phase. Since $v_\nu$ is the ``intrinsic'' waveform (not filtered through the detector response), we can apply the ordinary SPA to it and write it in a more general form as a function of the Fourier frequency as follows:
\beqa
v_F &=& A_F e^{i\psi_F}\,, \\
\psi_F &=& 2\pi\int\limits_0^{t(F)}{f(t')\,\dd{t'}} - 2\pi F\,t(F) \nonumber \\
&=& -2\pi\int\limits_{f_0}^{F}{t(f')\,\dd{f'}}\,,
\eeqa
whence
\beqa
v_{F-m/T} &\approx & A_F \exp(i\psi_{F-m/T}) \nonumber \\
&\approx& v_F \exp\left(im\frac{2\pi t(F)}{T}\right)\,, \\
\sum\limits_m{\widetilde{a}_m v_{F-m/T}} &\approx& v_F \sum\limits_m{\widetilde{a}_m \exp\left(im\frac{2\pi t(F)}{T}\right)} \nonumber \\
&=& v_F\left.\left(\mathcal{F}e^{i\psi_D}\right)\right|_{t=t(F)}\,. \label{eq:harmonicsToSPA}
\eeqa
That is, we get precisely the SPA prescription in which the time in the LISA Doppler and pattern factors is subsituted for $t(F)$, the inverse of $f=f(t)$\,. Note that the decomposition of the phase is not valid for the QMSs. For a source with a linear drift, the total change in the phase $\Delta\psi_F = -\pi \dot{f}_0 T_{\rm obs}^2 = - 2\pi N$ for $F\in[f_0,f_0+\dot{f}_0 T_{\rm obs}]$, while the correction $2\pi m t/T\sim 2\pi K$, which exceed $|\Delta\psi_F|$ as long as $N\lesssim K = 10$. 

In Fig.~\ref{fig:spa_LISA} we show the FD waveform (top panel, Eq.~(\ref{eq:quasi_convolved_waveform})) and its SPA version (bottom panel, Eq.~(\ref{eq:harmonicsToSPA})) for a higher frequency drift $\dot{f}_0=5\,\mbox{yr}^{-2}>1/T$. In order to mitigate potential numerical issues with rapid oscillations in the intrinsic waveform, Eq.~(\ref{eq:fdotFresnel}), we use its SPA counterpart
\beqa
v_{F}^{\rm spa} &=& \frac{e^{i\psi_F}}{K\sqrt{\dot{f}_0T^2}}\,, \quad \psi_F = \frac{\pi(F-f_0)^2}{\dot{f}_0} + \frac{\pi}{4}\,, \label{eq:main_FourierPhase}
\eeqa
when $F\in[f_0,f_0+\dot{f}_0 T_{\rm obs}]$, and $v_F=0$ otherwise (see Appendix~\ref{app:sec:FresnelSPA} for details of the transition). Also, the squared amplitude of the FD waveform (top panel) is smoothed out on a scale of~$\approx 1\,\mbox{yr}^{-1}$ to better illustrate the qualitative resemblance. Although there are noticeable differences between the waveform and its SPA (which are expected, due to the various approximations we have introduced), the similarity between the two is clearly visible.

\section{Conclusions\label{sec:conclusions}}

We have considered the applicability of the conventional SPA to QMSs, defined as sources that complete only a few extra cycles due to their frequency drift throughout the duration of the LISA mission.
Equivalently, for these sources the total change in frequency during the observation time is smaller than $1/T=1\,\mbox{yr}^{-1}$, the characteristic frequency of the LISA detector response.
The drift can be induced either by GW emission alone (for detached binaries) or by other processes such as mass transfer (in interacting binaries).

We have demonstrated in detail how copies of the FD waveform of a QMS appear at multiple frequencies $F=f_0+m/T$ ($m$ is an integer) and how this ``line splitting'' can affect Fisher matrix calculations. We find that, unless a GW source is located close to one of the poles relative to the plane of the ecliptic, the contribution of higher harmonics $|m|>1$ cannot be neglected, and it even dominates at modest to low ecliptic latitudes. That is, these harmonics must be taken into account when computing the SNR of the source or the components of the Fisher matrix. 

We have also studied the dependence of the ``line splitting'' on the four angles that specify the sky position of the source and the orientation of its orbital plane. We have demonstrated that it is the polar angle~$\overline{\theta}_S$ (i.e., the ecliptic latitude) that affects the magnitude of the effect the most. Since the Galactic DWDs concentrate (obviously) towards the Galactic plane and center, their ecliptic latitudes are moderate to low, and the higher harmonics must be taken into account in most cases of interest. The same applies to heavier compact binaries composed of either stellar-mass black holes or neutron stars, if they emit GWs at $f_0\lesssim 1\,\mbox{mHz}$.

As a by-product of this study, we make publicly available online \texttt{lisajous}~\cite{lisajous}, a code snippet that interactively generates closed contours (``Lissajous curves'') of the inclination-weighted LISA pattern functions in the complex plane and the respective Fourier harmonics in the Fourier plane. In Figs.~\ref{app:fig:LISAjousFigure} and~\ref{app:fig:FourierFigure} we show a few snapshots of the contours generated in this way and of their Fourier coefficients.

Let us briefly discuss some technical aspects of the calculation presented in this paper.

First, recall that we assumed $T_{\rm obs}=10\,\mbox{yr}$, while the LISA mission lifetime may be shorter (e.g., $4$ or $6$~yr~\cite{Seoane:2021kkk}). Shorter observation times do not change the qualitative picture: in fact, they would result into a larger number of QMSs, because the GW frequency drift is smaller for shorter observation times. By the same token, our results are applicable to lower-frequency detectors such as $\mu$Ares~\cite{Sesana:2019vho}, a proposed GW detector for $\mu$Hz frequencies whose detector response would also be periodic on a timescale of~$1.7\;\mbox{yr}$ (the orbital period of Mars).

Second, in our calculations we used a rectangular window to account for the finite length of the GW signal, whereas in practice windows that prevent spectral leakage (such as the Tukey window) are preferred. Qualitatively, this choice does not affect our results either. The specific choice of window only changes the shape of the functions $u_\nu$, Eq.~(\ref{eq:uFunctions}), but not the fact that they fall off on a scale of $\Delta F\sim 1/T_{\rm obs}$ (inverse window length).

Finally, we note that the decomposition of a QMS waveform in the FD domain given by Eq.~(\ref{eq:quasi_convolved_waveform}) is a particular case of the so-called fast-slow decomposition (see the Appendix of Ref.~\cite{Cornish:2007if}), and it appears to be quite convenient for the calculation of Fisher matrix uncertainties, either numerically or through autodifferentiation~\cite{jax2018github}. In each term the angular dependence is decoupled from the actual FD waveform and encoded in the coefficients~$\widetilde{a}_m$. These coefficients, as well as their derivatives with respect to the angular variables, are numerically well-behaved, and they can be evaluated rather quickly and robustly. Regarding the linear-drift FD waveform itself, Eq.~(\ref{eq:fdotFresnel}), it is given in terms of the Fresnel integrals, which are mathematically well studied and implemented in most scientific software packages. We leave an implementation of the decomposition~(\ref{eq:quasi_convolved_waveform}), either numerically or through autodifferentiation, and its application to Fisher parameter estimation studies, to future work.

\begin{acknowledgments}
We thank Valeriya Korol, Katelyn Breivik and Reza Ebadi for help and advice on the synthetic DWD populations, and Erwin Tanin for discussions on the effect of windowing. E.B. and V.S. are supported by NSF Grants No.~AST-2006538, PHY-2207502, PHY-090003 and PHY-20043, by NASA Grants No.~20-LPS20-0011 and~21-ATP21-0010, and by the John Templeton Foundation Grant~62840. E.B. and V.S. acknowledge support from the ITA-USA Science and Technology Cooperation program supported by the Ministry of Foreign Affairs of Italy (MAECI) and from the Indo-US Science and Technology Forum through the Indo-US Centre for Gravitational-Physics and Astronomy, grant~IUSSTF/JC-142/2019. This work was carried out at the Advanced Research Computing at Hopkins (ARCH) core facility (\url{rockfish.jhu.edu}), which is supported by the NSF Grant No.~OAC-1920103.
  
\textit{Software}. IPython~\citep{2007CSE.....9c..21P}, SciPy~\citep{2020NatMe..17..261V},  Matplotlib~\citep{2007CSE.....9...90H}, NumPy~\citep{2011CSE....13b..22V}, SymPy~\citep{Meurer:2017yhf}, \texttt{mpmath}~\citep{mpmath}, \texttt{filltex}~\citep{2017JOSS....2..222G}.
\end{acknowledgments}

\appendix

\section{A refresher on $L^2$ inner product\label{app:sec:inner}}

Since the $L^2$ product $\langle f,g\rangle = \int_\mathrm{R}{f g^*\,\dd t}$ is invariant under Fourier transform, for real-valued functions in the time domain and for a single data stream~$I$, 
\beqa
(\widetilde{h}^{(1)}|\widetilde{h}^{(2)})_{\rm I}&=&\langle\widetilde{h}^{(1)},\widetilde{h}^{(2)}\rangle = \langle h^{(1)}_F,h^{(2)}_F\rangle \nonumber \\
&=& \int\limits_0^{+\infty}{\frac{h^{(1)}_F\left(h^{(2)}_F\right)^* + h^{(1)}_{-F}\left(h^{(2)}_{-F}\right)^*}{S_{\rm n}(F)/2}\,\dd{F}} \nonumber \\
&=& 4\,{\rm Re}\int\limits_0^{+\infty}{\frac{h^{(1)}_F\left(h^{(2)}_F\right)^*}{S_{\rm n}(F)}\,\dd{F}} = (h^{(1)}_F\,|\,h^{(2)}_F)_{\rm I}\,, \nonumber \\
&{}& \label{app:eq:inner_def}
\eeqa
where the additional factor of $1/2$ for the noise spectrum is due to the definition of~$S_{\rm n}$ as a one-sided spectral density, and we have also used the property $h^*_{F} = h_{-F}$ for the Fourier transform of a real-valued function.

If $h_F$, or~$\widetilde{h}$, depends on a parameter~$\theta$ while $(h_F|h_F)$ does not, a useful property is that
\be
\left(h_F\left|\frac{\partial h_F}{\partial\theta}\right.\right) = 0\,. \label{app:eq:inner1}
\ee
This can be seen as follows:
\beqa
0 &=& \frac{\partial}{\partial\theta}(h_F|h_F) \nonumber \\
&=& \left(\left.\frac{\partial h_F}{\partial\theta}\right|h_F\right) + \left(h_F\left|\frac{\partial h_F}{\partial\theta}\right.\right) \nonumber \\
&=&\left(h_F\left|\frac{\partial h_F}{\partial\theta}\right.\right)^* + \left(h_F\left|\frac{\partial h_F}{\partial\theta}\right.\right) \nonumber \\
&=& 2\,{\rm Re}\left(h_F\left|\frac{\partial h_F}{\partial\theta}\right.\right) = 2\left(h_F\left|\frac{\partial h_F}{\partial\theta}\right.\right)\,, \label{app:eq:inner2}
\eeqa
where in the last step we omitted the real part, because the inner product is real in our case (see Eq.~(\ref{app:eq:inner_def})).

\section{SNR of a monochromatic source in the TD\label{app:sec:SNR}}

Using the definition of SNR, Eq.~(\ref{eq:SNR_TD_def}), and the TD waveform, Eq.~(\ref{eq:TD_waveform}), we obtain for a source with a constant frequency $f_0$: 
\beqa
&{}&\mbox{SNR}^2 = 2\int\limits_0^{T_{\rm obs}}{\dd{t}\,\frac{h^2(t)}{S_{\rm n}(f_0)}} \nonumber \\
&=& \frac{A^2}{S_{\rm n}(f_0)}\int\limits_0^{T_{\rm obs}}{\dd{t}\,\left[1+\cos{\left(4\pi f_0 t +2\psi_0 + 2\psi_D\right)}\right]} \nonumber \\
&\equiv& \frac{A^2T_{\rm obs}}{S_{\rm n}(f_0)}\left(1 + \Delta I\right)\,,
\eeqa
where
\begin{widetext}
\beqa
\Delta I &\equiv& \frac{1}{T_{\rm obs}}\int\limits_0^{T_{\rm obs}}{\dd{t}\,\cos{\left[4\pi f_0 t +2\psi_0 + 4\pi f_0\overline{R}\cos{\left(\overline{\phi}(t)-\overline{\phi}_S\right)}\right]}} \nonumber \\
&=& {\rm Re}\int\limits_0^{2\pi K}{\frac{\dd\overline{\phi}}{2\pi K}\,e^{2if_0 T\overline{\phi} + 2i\psi_0 + 4\pi i f_0\overline{R}\cos{\left(\overline{\phi}-\overline{\phi}_S\right)}} } = {\rm Re}\sum\limits_{m=-\infty}^{+\infty}{i^mJ_m\left(4\pi f_0\overline{R}\right) e^{2i\psi_0 - im\overline{\phi}_S}\int\limits_0^{2\pi K}{\frac{\dd{\overline{\phi}}}{2\pi K}\,e^{i\overline{\phi}(m+2f_0 T)}}} \nonumber \\
&=& {\rm Re}\sum\limits_{m=-\infty}^{+\infty}{J_m\left(4\pi f_0\overline{R}\right) e^{2i\psi_0 - im\overline{\phi}_S+im\frac{\pi}{2}}e^{i\pi K(m+2f_0 T)}\frac{\sin\left[\pi K(m+2f_0 T)\right]}{\pi K(m+2f_0 T)}} \nonumber \\
&=& \sum\limits_{m=-\infty}^{+\infty}{J_m\left(4\pi f_0\overline{R}\right) \cos{\left(2\pi f_0 T_{\rm obs} + 2\psi_0 - m\overline{\phi}_S+\frac{m\pi}{2}\right)}\frac{\sin\left[\pi K(m+2f_0 T)\right]}{\pi K(m+2f_0 T)}}\,.
\eeqa
\end{widetext}
Here we introduce the notation $K=T_{\rm obs}/T$ and we assume that $K$ is an even integer.

\begin{figure}[t]
    \includegraphics[width=0.85\columnwidth]{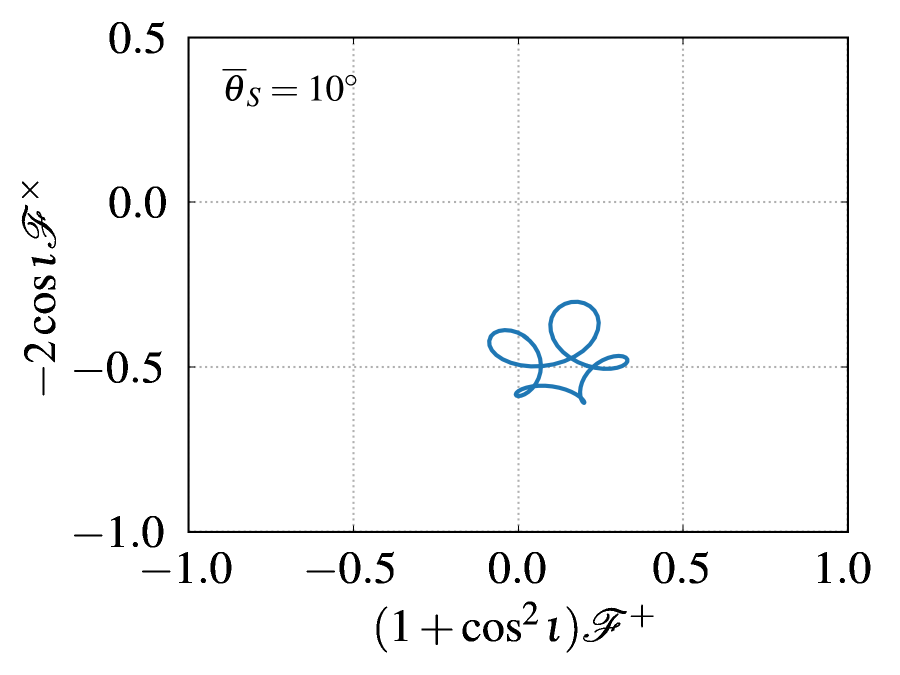}
    \includegraphics[width=0.85\columnwidth]{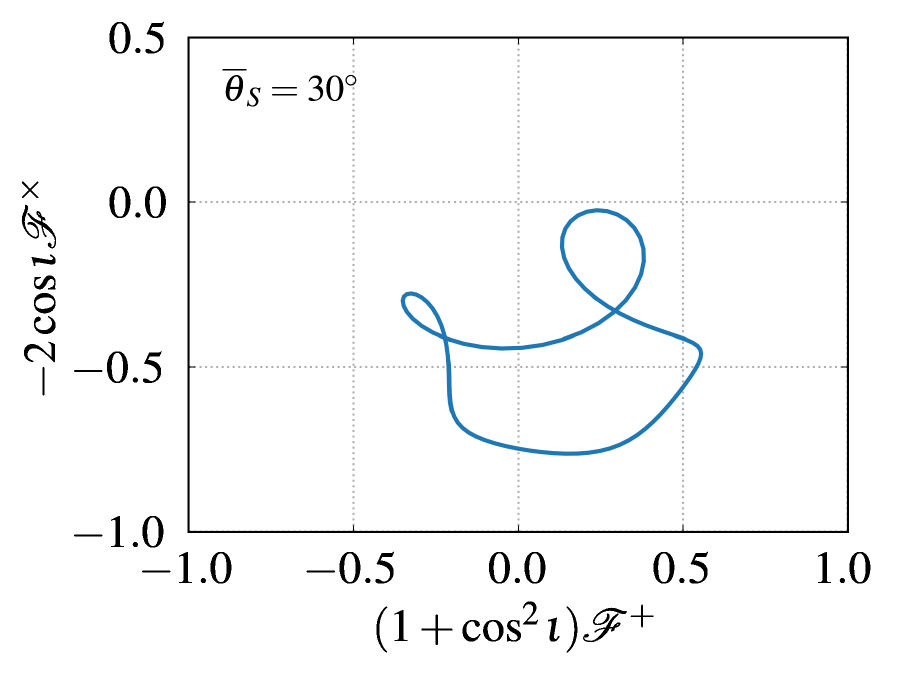}
    \includegraphics[width=0.85\columnwidth]{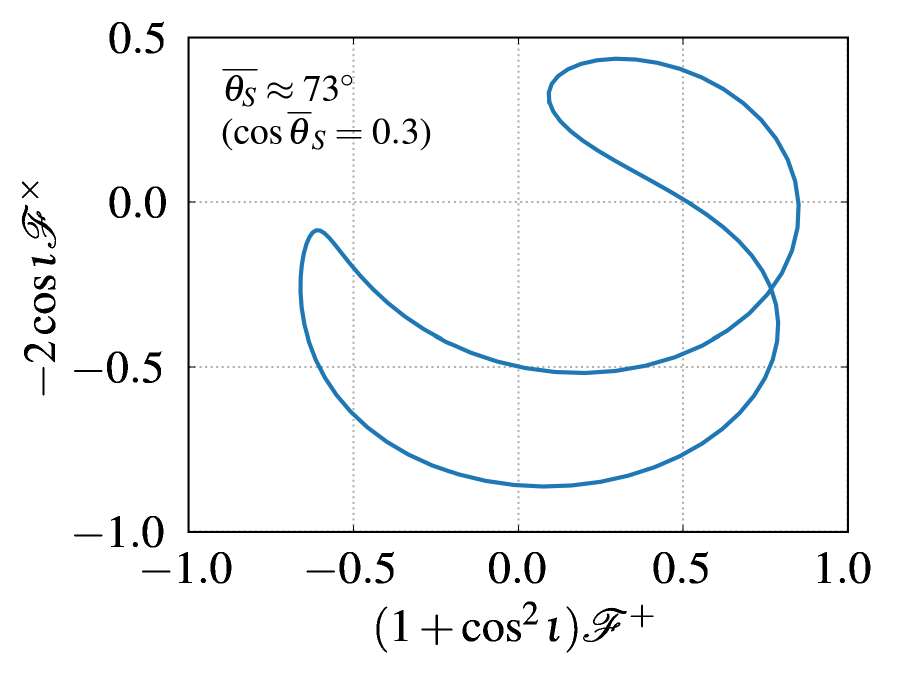}
    \caption{Closed contours (``Lissajous curves'') traced by the LISA pattern function factor~$\mathcal{F}$ in the complex plane, with the inclination-weighted plus and cross polarization patterns plotted as real and imaginary parts, respectively: see Eq.~(\ref{eq:patternTD}). The source angle~$\overline{\theta}_S$ (i.e., the angular distance from the ecliptic pole) increases from top to bottom: $\overline{\theta}_S=10^\circ$ (top), $\overline{\theta}_S=30^\circ$ (middle), and $\overline{\theta}_S=\arccos{0.3}$ (bottom). The other angles are fixed at their fiducial values: $\overline{\phi}_S=5$, $\overline{\theta}_L=\arccos{(-0.2)}$, and $\overline{\phi}_L=4$.
    \label{app:fig:LISAjousFigure}}%
\end{figure}

Now, note that, for the applications under consideration, $f_0 T\gg 1$. In the sum above, there are essentially two factors that contribute to the amplitude of each term: the Bessel function and the sinc function $\sinc{y}\equiv\sin y/y$. The sinc function is of order unity when $|m|\approx 2f_0 T\gg 1$, which implies
\beqa
J_m(x) &\approx& \frac{1}{\sqrt{2\pi m}}\left(\frac{ex}{2m}\right)^m \ll 1\,,\\
&{}& \qquad x\sim 2\pi f_0R\sim 10\,,
\eeqa
where the Bessel function at large~$m$ falls off very rapidly. On the other hand, $J_m(x)\sim 1$ when $m\sim 1$, which results in the sinc function $\sim 1/(f_0 T)$, so that
\be
\Delta I = \mathcal{O}\left[(f_0 T)^{-1}\right]\,.
\ee

\section{LISA Pattern Functions\label{app:sec:patternNumerical}}

The LISA pattern function factor~$\mathcal{F}$, defined in Eq.~(\ref{eq:patternTD}), is a complex-valued function of time with period $T=1\,\mbox{yr}$. Hence it traces a closed contour in the complex plane. Here we provide a few examples of the contours. We also show contours that the Fourier coefficients of~$\mathcal{F}$ follow when all angles but one are kept fixed. Recall that we use the fiducial values~\citep{Takahashi:2002ky} $\overline{\theta}_S=\arccos{0.3}$, $\overline{\phi}_S=5$, $\overline{\theta}_L=\arccos{(-0.2)}$, and $\overline{\phi}_L=4$ (the second pair of angles is converted to the inclination~$\iota$ and polarization angles~$\varphi$).

To illustrate the behavior of the LISA pattern function contours and of their Fourier transforms for any combinations of the angles, we provide \texttt{lisajous}
(LISA + ``Lissajous curves'')~\cite{lisajous}, a code for interactive plots with sliders.

In Fig.~\ref{app:fig:LISAjousFigure} we show the complex plane of~$\mathcal{F}$, where the real and imaginary parts are the inclination-weighted plus and cross polarization patterns: $(1+\cos^{2}{\iota})\mathcal{F}^{+}$ and $-2\cos\iota\,\mathcal{F}^{\times}$, respectively. The three panels correspond to different values of the angle~$\overline{\theta}_S$: $10^\circ$ (top), $30^\circ$ (middle), and $\arccos{0.3}\approx 73^\circ$ (bottom). In the limit $\overline{\theta}_S=0^\circ$ or $\overline{\theta}_S=180^\circ$, the contour is degenerate and reduces to a point. 

\begin{figure*}[!htbp]
    \includegraphics[width=0.9\columnwidth]{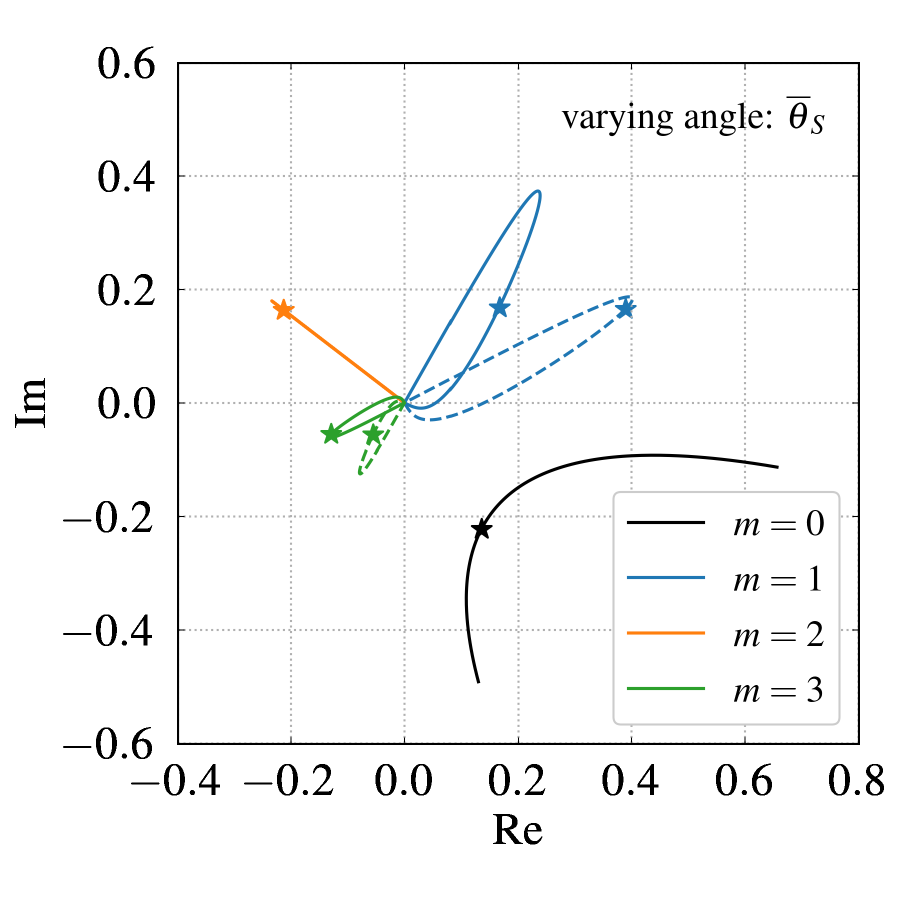}
    \includegraphics[width=0.9\columnwidth]{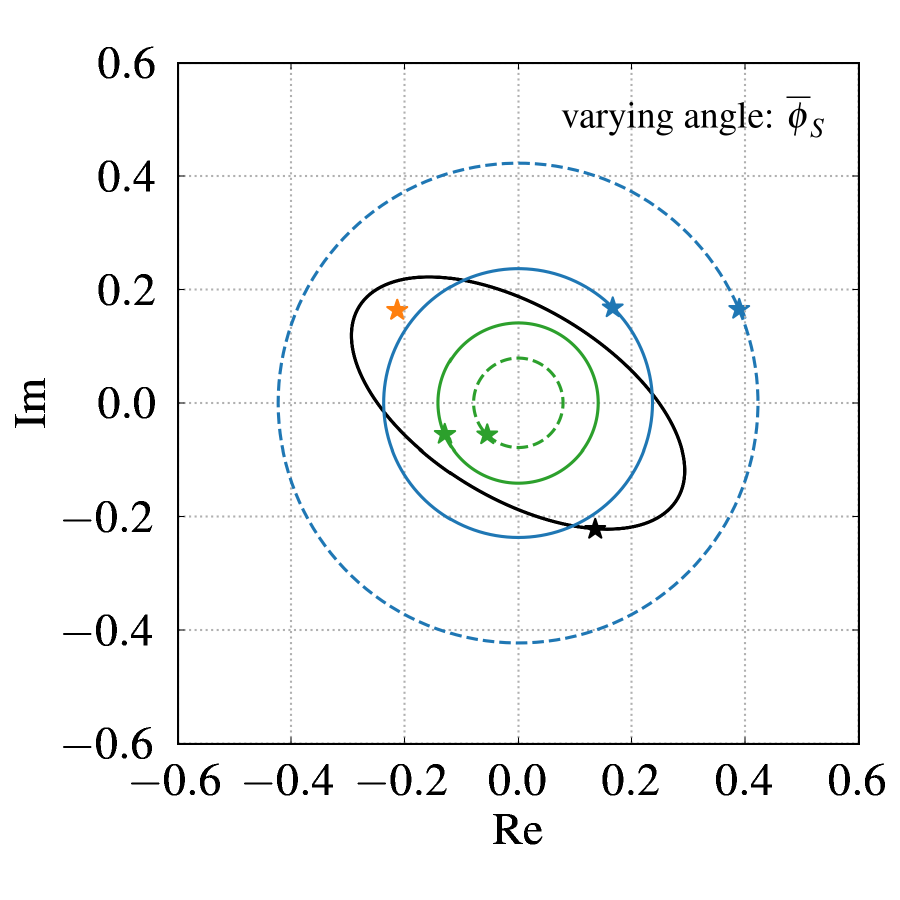}
    \includegraphics[width=0.9\columnwidth]{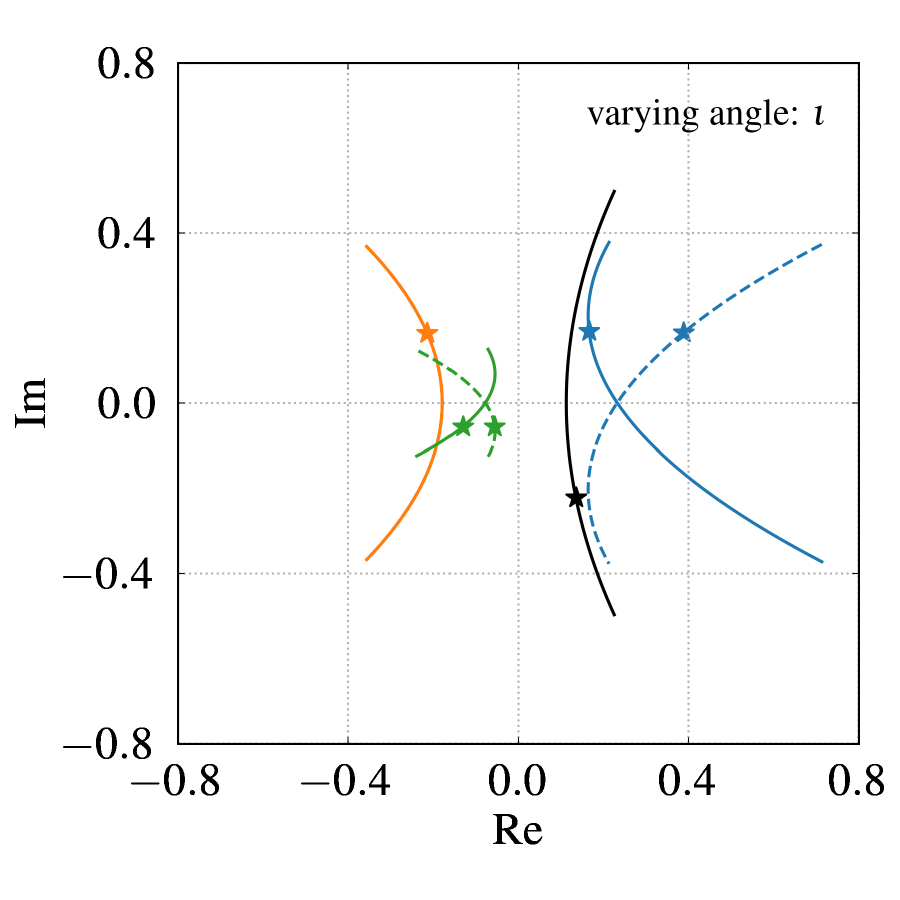}
    \includegraphics[width=0.9\columnwidth]{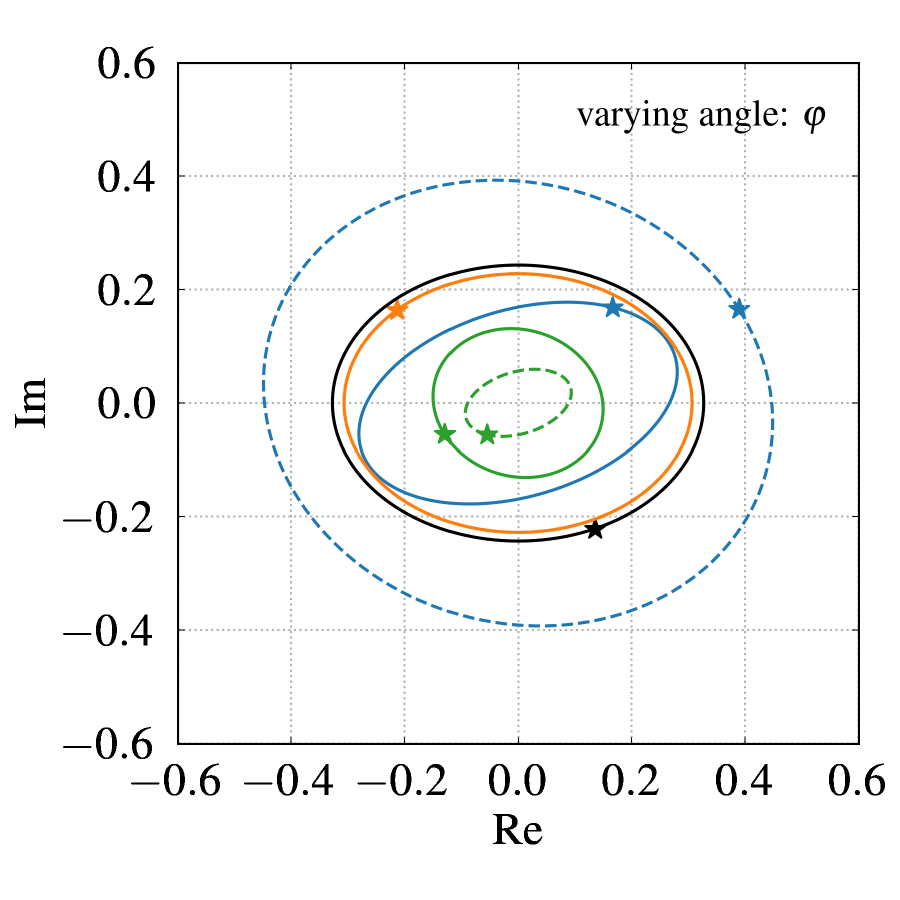}
    \caption{Fourier harmonics $m=0,\pm 1,\pm 2, \pm 3$ of the LISA pattern factor~$\mathcal{F}$, Eqs.~(\ref{eq:patternTD})--(\ref{eq:patternFourier}), as a function of the angles $\overline{\theta}_S$, $\overline{\phi}_S$, $\iota$, and $\varphi$. Each panel corresponds to the case in which one of the angles, indicated in the legend, starts from its fiducial value (marked with a star) and varies, while the other angles are fixed to their fiducial values: $\overline{\theta}_S=\arccos{0.3}\approx 73^\circ$, $\overline{\phi}_S=5$, $\iota\approx 64^\circ$, $\varphi\approx 113^\circ$. For $m\neq 0$, solid and dashed curves of the same color depict the positive and negative coefficients $\pm |m|$, respectively. The double-frequency harmonic coincides with its negative counterpart (for the arm~I case shown here).
    \label{app:fig:FourierFigure}}%
\end{figure*}

In Fig.~\ref{app:fig:FourierFigure} we show the complex planes of the Fourier harmonics $m=0,\pm 1,\pm 2,\pm 3$ of the pattern function. Each panel corresponds to the paths followed by the Fourier coefficients when one of the angles (indicated in the legend) varies, while the others are fixed to their fiducial values. The black solid line depicts the constant harmonic $m=0$, whereas the solid/dashed pairs of the same color are the paths for the positive/negative coefficient of the same order~$m$. There is only a solid line for the double-frequency harmonic, because the paths of $m=2$ and $m=-2$ coincide (for arm~I of the detector). The starred dots on each curve mark the fiducial value of the varying angle. Overall, one can see that the amplitudes of all the Fourier coefficients are different from zero except for the case of varying~$\overline{\theta}_S$, where the harmonics with $m\neq 0$ vanish when $\overline{\theta}_S=0^\circ, 180^\circ$. 

On a final note, one way to interpret the Fourier series for the LISA pattern function is view it as a superposition of elliptically polarized waves with frequencies that are multiples of~$1/T$. (This is, of course, only a helpful interpretation, since the frequency~$1/T$ is beyond LISA's frequency band). Proportions of the ellipse are defined by the relative magnitudes and phases of the complex amplitudes
\beqa
h^{(m)}_{+} &=& (1+\cos^{2}\iota)b^{+}_m\,, \\
h^{(m)}_{\times} &=& -2i\cos\iota\,b^{\times}_m\,.
\eeqa

\section{SPA for a waveform with linear frequency drift\label{app:sec:FresnelSPA}}

\begin{figure*}[!htbp]
    \includegraphics[width=0.32\linewidth]{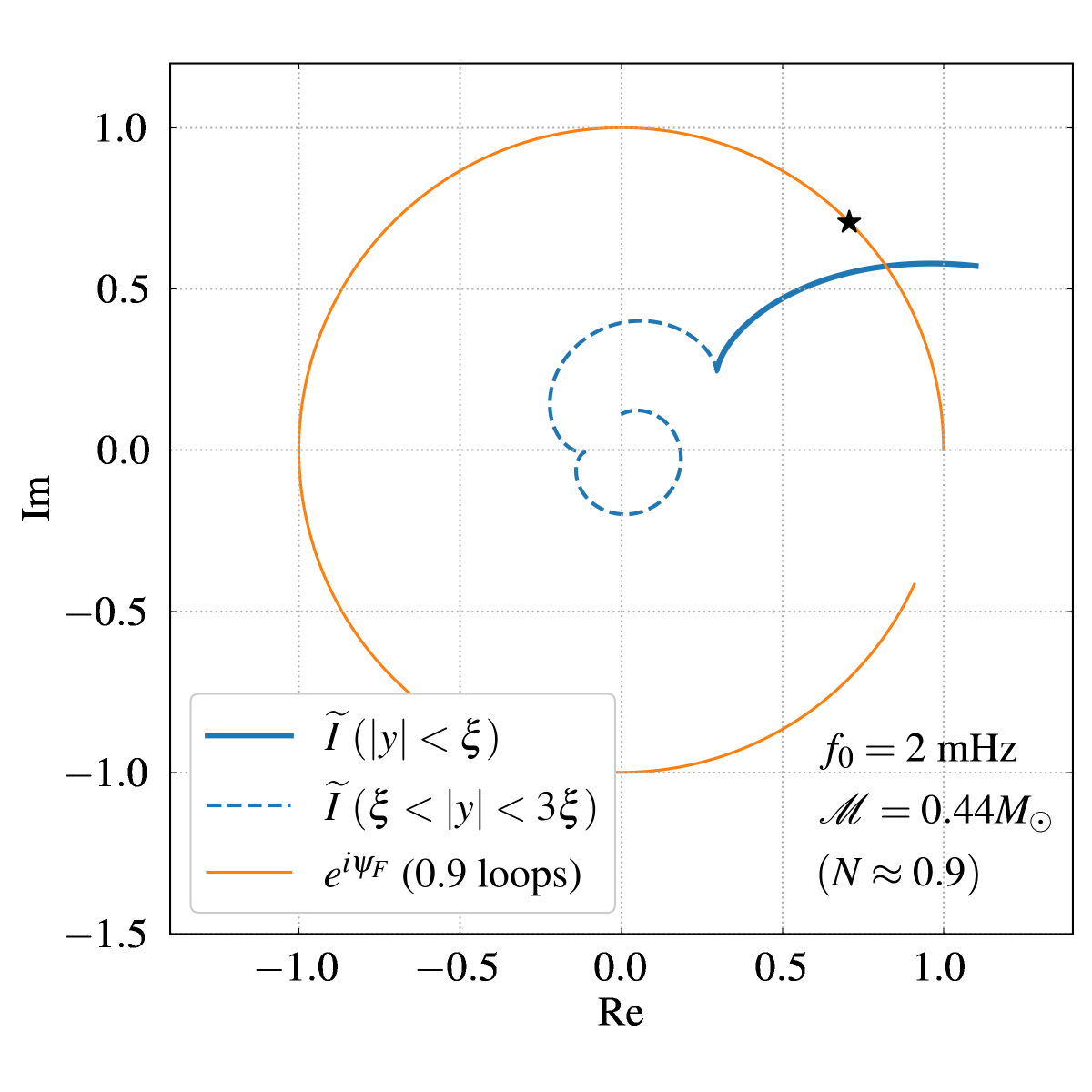}
    \includegraphics[width=0.32\linewidth]{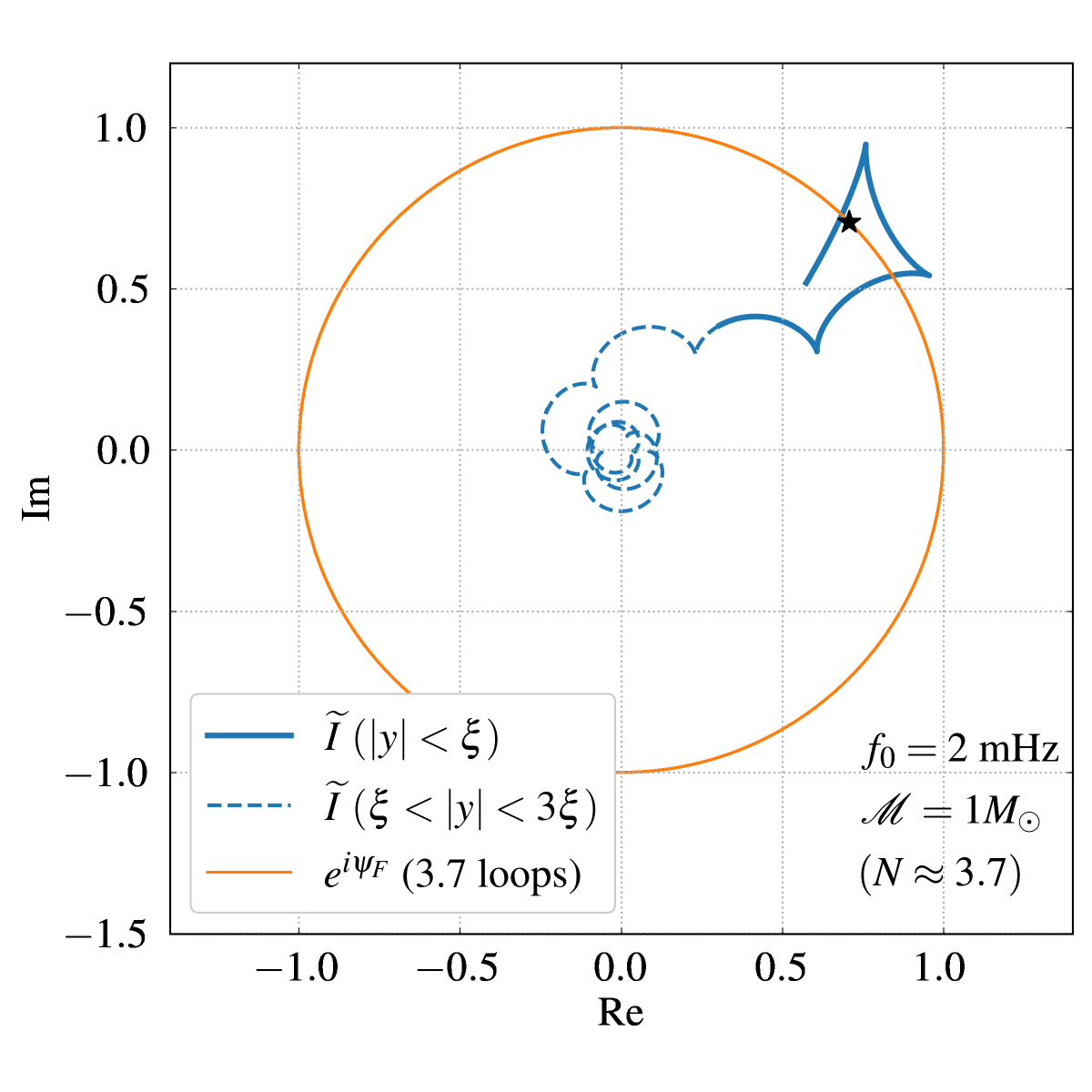}
    \includegraphics[width=0.32\linewidth]{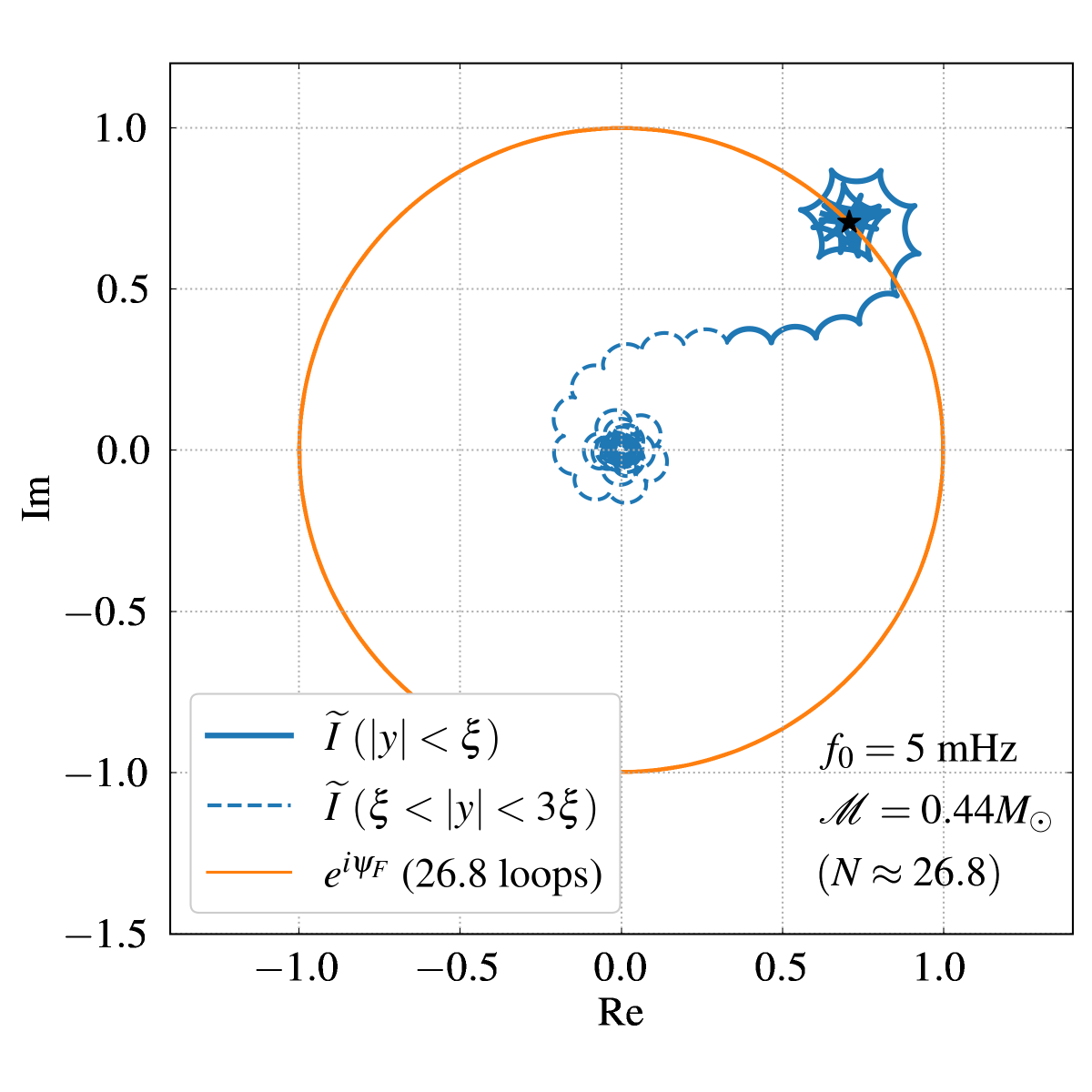}
    \caption{Comparison between contributions to the FD waveform of a source with linear frequency drift from the main Fourier phase (orange line, shifted to $\psi_F=0$), Eq.~(\ref{eq:main_FourierPhase}), and from the prefactor in Eq.~(\ref{eq:fdotFresnel}) normalized to its SPA value, $\widetilde{I}\equiv I/|I_0|$. The main phase contribution is shown for a range of $t\in[0,T_{\rm obs}]$ ($|y|\leq\xi=\sqrt{N}$ in dimensionless variables), which corresponds to Fourier frequencies $F\in[f_0,f_0+\dot{f}_0 T_{\rm obs}]$\,. The total number of loops of the orange curve is indicated in the legend. The prefactor is shown for $|y|\leq\xi$ ($t\in[0,T_{\rm obs}]$, solid line) and for  $\xi\leq|y|\leq 3\xi$ ($t\in[T_{\rm obs},2T_{\rm obs}]$, dashed line), and the black star indicates the limit as the integration domain is extended to the entire real axis. As the dimensionless parameter $\xi=\sqrt{N}$ grows, the integral exhibits the behavior that is usually assumed in SPA calculations. The values of the parameters, of the chirp mass, and of the GW frequency are given in the legends, and the observation time $T_{\rm obs}=10$~yr for all panels.}
    \label{app:fig:SPAcorrection}
\end{figure*}

Since the windowed FD waveform can be written analytically in terms of the Fresnel integrals (see Eq.~(\ref{eq:fdotFresnel})), the case of the linear drift gives us an opportunity to better understand the relation between the exact FD waveform and its SPA counterpart.

The range of GW frequencies exhibited by a source during the observation period corresponds to $0\leq t\leq T_{\rm obs}$, or $-\xi\leq y\leq \xi$. For $\xi\gg 1$, the prefactor in Eq.~(\ref{eq:fdotFresnel}), which we denote by $I$ here, has its SPA value
\be
I=I_0 \approx \frac{2Z(\xi)}{K\sqrt{2\dot{f}_0 T^2}} \approx \frac{e^{\frac{i\pi}{4}}}{K\sqrt{\dot{f}_0 T^2}}\,,
\label{app:eq:I0}
\ee
where we have used the fact that $Z(+\infty)=-Z(-\infty)=\frac{1}{\sqrt{2}}e^{\frac{i\pi}{4}}$. At the ends of the interval $y=\pm\xi$, the amplitude is smaller by a factor of two, $I\approx I_0/2$. Far beyond the interval, where $|y\pm \xi|\gtrsim \xi$, the integral vanishes, because $Z(\xi+y)+Z(\xi-y)\approx Z(y)+Z(-y)=0$\,.

However, for the QMSs under consideration the approximation~$\xi\gg 1$ is barely satisfied, because \mbox{$\xi=\sqrt{N}\lessapprox 3$}.

In Fig.~\ref{app:fig:SPAcorrection} we compare the prefactor~$\widetilde{I}\equiv I/|I_0|$ (blue lines) normalized by the SPA amplitude~$|I_0|$ to its SPA value (marked with a black star). We do so for different values of the number of extra cycles that increase from left to right: $N\approx 0.9$ (left panel), $N\approx 3.7$ (middle panel), and $N\approx 26.8$ (right panel). These values correspond to specific combinations of the GW frequency~$f_0$ and chirp mass~$\mathcal{M}$ (see the legend). We also show the winding of the main Fourier phase (orange line, see also Eq.~(\ref{eq:main_FourierPhase})) and indicate the number of loops in the legend. The normalized prefactor~$\widetilde{I}$ is shown for two ranges of the frequencies: the conventional SPA range $0<(F-f_0)<\dot{f}_0 T_{\rm obs}$ (solid line) and a range extended by $(-\dot{f}_0 T_{\rm obs})$ on each side (dashed line). Recall that, in the SPA, the FD waveform is assumed to quickly fall off outside of the SPA range. 

\newpage

\bibliography{refs}%

\begin{thebibliography}{42}%
\makeatletter
\providecommand \@ifxundefined [1]{%
 \@ifx{#1\undefined}
}%
\providecommand \@ifnum [1]{%
 \ifnum #1\expandafter \@firstoftwo
 \else \expandafter \@secondoftwo
 \fi
}%
\providecommand \@ifx [1]{%
 \ifx #1\expandafter \@firstoftwo
 \else \expandafter \@secondoftwo
 \fi
}%
\providecommand \natexlab [1]{#1}%
\providecommand \enquote  [1]{``#1''}%
\providecommand \bibnamefont  [1]{#1}%
\providecommand \bibfnamefont [1]{#1}%
\providecommand \citenamefont [1]{#1}%
\providecommand \href@noop [0]{\@secondoftwo}%
\providecommand \href [0]{\begingroup \@sanitize@url \@href}%
\providecommand \@href[1]{\@@startlink{#1}\@@href}%
\providecommand \@@href[1]{\endgroup#1\@@endlink}%
\providecommand \@sanitize@url [0]{\catcode `\\12\catcode `\$12\catcode
  `\&12\catcode `\#12\catcode `\^12\catcode `\_12\catcode `\%12\relax}%
\providecommand \@@startlink[1]{}%
\providecommand \@@endlink[0]{}%
\providecommand \url  [0]{\begingroup\@sanitize@url \@url }%
\providecommand \@url [1]{\endgroup\@href {#1}{\urlprefix }}%
\providecommand \urlprefix  [0]{URL }%
\providecommand \Eprint [0]{\href }%
\providecommand \doibase [0]{http://dx.doi.org/}%
\providecommand \selectlanguage [0]{\@gobble}%
\providecommand \bibinfo  [0]{\@secondoftwo}%
\providecommand \bibfield  [0]{\@secondoftwo}%
\providecommand \translation [1]{[#1]}%
\providecommand \BibitemOpen [0]{}%
\providecommand \bibitemStop [0]{}%
\providecommand \bibitemNoStop [0]{.\EOS\space}%
\providecommand \EOS [0]{\spacefactor3000\relax}%
\providecommand \BibitemShut  [1]{\csname bibitem#1\endcsname}%
\let\auto@bib@innerbib\@empty
%</preamble>
\bibitem [{\citenamefont {{Amaro Seoane}}\ \emph {et~al.}(2017)\citenamefont
  {{Amaro Seoane}} \emph {et~al.}}]{2017arXiv170200786A}%
  \BibitemOpen
  \bibfield  {author} {\bibinfo {author} {\bibfnamefont {P.}~\bibnamefont
  {{Amaro Seoane}}} \emph {et~al.},\ }\href {\doibase
  10.48550/arXiv.1702.00786} {\bibfield  {journal} {\bibinfo  {journal} {arXiv
  e-prints}\ ,\ \bibinfo {eid} {arXiv:1702.00786}} (\bibinfo {year} {2017})},\
  \Eprint {http://arxiv.org/abs/1702.00786} {arXiv:1702.00786 [astro-ph.IM]}
  \BibitemShut {NoStop}%
\bibitem [{\citenamefont {Seoane}\ \emph {et~al.}(2022)\citenamefont {Seoane}
  \emph {et~al.}}]{Seoane:2021kkk}%
  \BibitemOpen
  \bibfield  {author} {\bibinfo {author} {\bibfnamefont {P.~A.}\ \bibnamefont
  {Seoane}} \emph {et~al.},\ }\href {\doibase 10.1007/s10714-021-02889-x}
  {\bibfield  {journal} {\bibinfo  {journal} {Gen. Rel. Grav.}\ }\textbf
  {\bibinfo {volume} {54}},\ \bibinfo {pages} {3} (\bibinfo {year} {2022})},\
  \Eprint {http://arxiv.org/abs/2107.09665} {arXiv:2107.09665 [astro-ph.IM]}
  \BibitemShut {NoStop}%
\bibitem [{\citenamefont {Amaro~Seoane}\ \emph {et~al.}(2023)\citenamefont
  {Amaro~Seoane} \emph {et~al.}}]{LISA:2022yao}%
  \BibitemOpen
  \bibfield  {author} {\bibinfo {author} {\bibfnamefont {P.}~\bibnamefont
  {Amaro~Seoane}} \emph {et~al.} (\bibinfo {collaboration} {LISA}),\ }\href
  {\doibase 10.1007/s41114-022-00041-y} {\bibfield  {journal} {\bibinfo
  {journal} {Living Rev. Rel.}\ }\textbf {\bibinfo {volume} {26}},\ \bibinfo
  {pages} {2} (\bibinfo {year} {2023})},\ \Eprint
  {http://arxiv.org/abs/2203.06016} {arXiv:2203.06016 [gr-qc]} \BibitemShut
  {NoStop}%
\bibitem [{\citenamefont {Abbott}\ \emph {et~al.}(2023)\citenamefont {Abbott}
  \emph {et~al.}}]{KAGRA:2021duu}%
  \BibitemOpen
  \bibfield  {author} {\bibinfo {author} {\bibfnamefont {R.}~\bibnamefont
  {Abbott}} \emph {et~al.} (\bibinfo {collaboration} {KAGRA, VIRGO, LIGO
  Scientific}),\ }\href {\doibase 10.1103/PhysRevX.13.011048} {\bibfield
  {journal} {\bibinfo  {journal} {Phys. Rev. X}\ }\textbf {\bibinfo {volume}
  {13}},\ \bibinfo {pages} {011048} (\bibinfo {year} {2023})},\ \Eprint
  {http://arxiv.org/abs/2111.03634} {arXiv:2111.03634 [astro-ph.HE]}
  \BibitemShut {NoStop}%
\bibitem [{\citenamefont {Evans}\ \emph {et~al.}(2021)\citenamefont {Evans}
  \emph {et~al.}}]{Evans:2021gyd}%
  \BibitemOpen
  \bibfield  {author} {\bibinfo {author} {\bibfnamefont {M.}~\bibnamefont
  {Evans}} \emph {et~al.},\ }\href@noop {} {\  (\bibinfo {year} {2021})},\
  \Eprint {http://arxiv.org/abs/2109.09882} {arXiv:2109.09882 [astro-ph.IM]}
  \BibitemShut {NoStop}%
\bibitem [{\citenamefont {Maggiore}\ \emph {et~al.}(2020)\citenamefont
  {Maggiore} \emph {et~al.}}]{Maggiore:2019uih}%
  \BibitemOpen
  \bibfield  {author} {\bibinfo {author} {\bibfnamefont {M.}~\bibnamefont
  {Maggiore}} \emph {et~al.},\ }\href {\doibase 10.1088/1475-7516/2020/03/050}
  {\bibfield  {journal} {\bibinfo  {journal} {JCAP}\ }\textbf {\bibinfo
  {volume} {03}},\ \bibinfo {pages} {050} (\bibinfo {year} {2020})},\ \Eprint
  {http://arxiv.org/abs/1912.02622} {arXiv:1912.02622 [astro-ph.CO]}
  \BibitemShut {NoStop}%
\bibitem [{\citenamefont {Agazie}\ \emph {et~al.}(2023)\citenamefont {Agazie}
  \emph {et~al.}}]{NANOGrav:2023gor}%
  \BibitemOpen
  \bibfield  {author} {\bibinfo {author} {\bibfnamefont {G.}~\bibnamefont
  {Agazie}} \emph {et~al.} (\bibinfo {collaboration} {NANOGrav}),\ }\href
  {\doibase 10.3847/2041-8213/acdac6} {\bibfield  {journal} {\bibinfo
  {journal} {Astrophys. J. Lett.}\ }\textbf {\bibinfo {volume} {951}},\
  \bibinfo {pages} {L8} (\bibinfo {year} {2023})},\ \Eprint
  {http://arxiv.org/abs/2306.16213} {arXiv:2306.16213 [astro-ph.HE]}
  \BibitemShut {NoStop}%
\bibitem [{\citenamefont {Antoniadis}\ \emph {et~al.}(2023)\citenamefont
  {Antoniadis} \emph {et~al.}}]{EPTA:2023fyk}%
  \BibitemOpen
  \bibfield  {author} {\bibinfo {author} {\bibfnamefont {J.}~\bibnamefont
  {Antoniadis}} \emph {et~al.} (\bibinfo {collaboration} {EPTA}),\ }\href
  {\doibase 10.1051/0004-6361/202346844} {\bibfield  {journal} {\bibinfo
  {journal} {Astron. Astrophys.}\ }\textbf {\bibinfo {volume} {678}},\ \bibinfo
  {pages} {A50} (\bibinfo {year} {2023})},\ \Eprint
  {http://arxiv.org/abs/2306.16214} {arXiv:2306.16214 [astro-ph.HE]}
  \BibitemShut {NoStop}%
\bibitem [{\citenamefont {Reardon}\ \emph {et~al.}(2023)\citenamefont {Reardon}
  \emph {et~al.}}]{Reardon:2023gzh}%
  \BibitemOpen
  \bibfield  {author} {\bibinfo {author} {\bibfnamefont {D.~J.}\ \bibnamefont
  {Reardon}} \emph {et~al.},\ }\href {\doibase 10.3847/2041-8213/acdd02}
  {\bibfield  {journal} {\bibinfo  {journal} {Astrophys. J. Lett.}\ }\textbf
  {\bibinfo {volume} {951}},\ \bibinfo {pages} {L6} (\bibinfo {year} {2023})},\
  \Eprint {http://arxiv.org/abs/2306.16215} {arXiv:2306.16215 [astro-ph.HE]}
  \BibitemShut {NoStop}%
\bibitem [{\citenamefont {Xu}\ \emph {et~al.}(2023)\citenamefont {Xu} \emph
  {et~al.}}]{Xu:2023wog}%
  \BibitemOpen
  \bibfield  {author} {\bibinfo {author} {\bibfnamefont {H.}~\bibnamefont {Xu}}
  \emph {et~al.},\ }\href {\doibase 10.1088/1674-4527/acdfa5} {\bibfield
  {journal} {\bibinfo  {journal} {Res. Astron. Astrophys.}\ }\textbf {\bibinfo
  {volume} {23}},\ \bibinfo {pages} {075024} (\bibinfo {year} {2023})},\
  \Eprint {http://arxiv.org/abs/2306.16216} {arXiv:2306.16216 [astro-ph.HE]}
  \BibitemShut {NoStop}%
\bibitem [{\citenamefont {Korol}\ \emph {et~al.}(2022)\citenamefont {Korol},
  \citenamefont {Hallakoun}, \citenamefont {Toonen},\ and\ \citenamefont
  {Karnesis}}]{Korol:2021pun}%
  \BibitemOpen
  \bibfield  {author} {\bibinfo {author} {\bibfnamefont {V.}~\bibnamefont
  {Korol}}, \bibinfo {author} {\bibfnamefont {N.}~\bibnamefont {Hallakoun}},
  \bibinfo {author} {\bibfnamefont {S.}~\bibnamefont {Toonen}}, \ and\ \bibinfo
  {author} {\bibfnamefont {N.}~\bibnamefont {Karnesis}},\ }\href {\doibase
  10.1093/mnras/stac415} {\bibfield  {journal} {\bibinfo  {journal} {Mon. Not.
  Roy. Astron. Soc.}\ }\textbf {\bibinfo {volume} {511}},\ \bibinfo {pages}
  {5936} (\bibinfo {year} {2022})},\ \Eprint {http://arxiv.org/abs/2109.10972}
  {arXiv:2109.10972 [astro-ph.HE]} \BibitemShut {NoStop}%
\bibitem [{\citenamefont {{Babak}}\ \emph {et~al.}(2021)\citenamefont
  {{Babak}}, \citenamefont {{Hewitson}},\ and\ \citenamefont
  {{Petiteau}}}]{2021arXiv210801167B}%
  \BibitemOpen
  \bibfield  {author} {\bibinfo {author} {\bibfnamefont {S.}~\bibnamefont
  {{Babak}}}, \bibinfo {author} {\bibfnamefont {M.}~\bibnamefont {{Hewitson}}},
  \ and\ \bibinfo {author} {\bibfnamefont {A.}~\bibnamefont {{Petiteau}}},\
  }\href {\doibase 10.48550/arXiv.2108.01167} {\bibfield  {journal} {\bibinfo
  {journal} {arXiv e-prints}\ ,\ \bibinfo {eid} {arXiv:2108.01167}} (\bibinfo
  {year} {2021})},\ \Eprint {http://arxiv.org/abs/2108.01167} {arXiv:2108.01167
  [astro-ph.IM]} \BibitemShut {NoStop}%
\bibitem [{\citenamefont {Nelemans}\ \emph {et~al.}(2001)\citenamefont
  {Nelemans}, \citenamefont {Yungelson},\ and\ \citenamefont
  {Portegies~Zwart}}]{Nelemans:2001hp}%
  \BibitemOpen
  \bibfield  {author} {\bibinfo {author} {\bibfnamefont {G.}~\bibnamefont
  {Nelemans}}, \bibinfo {author} {\bibfnamefont {L.~R.}\ \bibnamefont
  {Yungelson}}, \ and\ \bibinfo {author} {\bibfnamefont {S.~F.}\ \bibnamefont
  {Portegies~Zwart}},\ }\href {\doibase 10.1051/0004-6361:20010683} {\bibfield
  {journal} {\bibinfo  {journal} {Astron. Astrophys.}\ }\textbf {\bibinfo
  {volume} {375}},\ \bibinfo {pages} {890} (\bibinfo {year} {2001})},\ \Eprint
  {http://arxiv.org/abs/astro-ph/0105221} {arXiv:astro-ph/0105221} \BibitemShut
  {NoStop}%
\bibitem [{\citenamefont {Korol}\ \emph {et~al.}(2017)\citenamefont {Korol},
  \citenamefont {Rossi}, \citenamefont {Groot}, \citenamefont {Nelemans},
  \citenamefont {Toonen},\ and\ \citenamefont {Brown}}]{Korol:2017qcx}%
  \BibitemOpen
  \bibfield  {author} {\bibinfo {author} {\bibfnamefont {V.}~\bibnamefont
  {Korol}}, \bibinfo {author} {\bibfnamefont {E.~M.}\ \bibnamefont {Rossi}},
  \bibinfo {author} {\bibfnamefont {P.~J.}\ \bibnamefont {Groot}}, \bibinfo
  {author} {\bibfnamefont {G.}~\bibnamefont {Nelemans}}, \bibinfo {author}
  {\bibfnamefont {S.}~\bibnamefont {Toonen}}, \ and\ \bibinfo {author}
  {\bibfnamefont {A.~G.~A.}\ \bibnamefont {Brown}},\ }\href {\doibase
  10.1093/mnras/stx1285} {\bibfield  {journal} {\bibinfo  {journal} {Mon. Not.
  Roy. Astron. Soc.}\ }\textbf {\bibinfo {volume} {470}},\ \bibinfo {pages}
  {1894} (\bibinfo {year} {2017})},\ \Eprint {http://arxiv.org/abs/1703.02555}
  {arXiv:1703.02555 [astro-ph.HE]} \BibitemShut {NoStop}%
\bibitem [{\citenamefont {Stroeer}\ and\ \citenamefont
  {Vecchio}(2006)}]{Stroeer:2006rx}%
  \BibitemOpen
  \bibfield  {author} {\bibinfo {author} {\bibfnamefont {A.}~\bibnamefont
  {Stroeer}}\ and\ \bibinfo {author} {\bibfnamefont {A.}~\bibnamefont
  {Vecchio}},\ }\href {\doibase 10.1088/0264-9381/23/19/S19} {\bibfield
  {journal} {\bibinfo  {journal} {Class. Quant. Grav.}\ }\textbf {\bibinfo
  {volume} {23}},\ \bibinfo {pages} {S809} (\bibinfo {year} {2006})},\ \Eprint
  {http://arxiv.org/abs/astro-ph/0605227} {arXiv:astro-ph/0605227} \BibitemShut
  {NoStop}%
\bibitem [{\citenamefont {Kupfer}\ \emph {et~al.}(2018)\citenamefont {Kupfer},
  \citenamefont {Korol}, \citenamefont {Shah}, \citenamefont {Nelemans},
  \citenamefont {Marsh}, \citenamefont {Ramsay}, \citenamefont {Groot},
  \citenamefont {Steeghs},\ and\ \citenamefont {Rossi}}]{Kupfer:2018jee}%
  \BibitemOpen
  \bibfield  {author} {\bibinfo {author} {\bibfnamefont {T.}~\bibnamefont
  {Kupfer}}, \bibinfo {author} {\bibfnamefont {V.}~\bibnamefont {Korol}},
  \bibinfo {author} {\bibfnamefont {S.}~\bibnamefont {Shah}}, \bibinfo {author}
  {\bibfnamefont {G.}~\bibnamefont {Nelemans}}, \bibinfo {author}
  {\bibfnamefont {T.~R.}\ \bibnamefont {Marsh}}, \bibinfo {author}
  {\bibfnamefont {G.}~\bibnamefont {Ramsay}}, \bibinfo {author} {\bibfnamefont
  {P.~J.}\ \bibnamefont {Groot}}, \bibinfo {author} {\bibfnamefont {D.~T.~H.}\
  \bibnamefont {Steeghs}}, \ and\ \bibinfo {author} {\bibfnamefont {E.~M.}\
  \bibnamefont {Rossi}},\ }\href {\doibase 10.1093/mnras/sty1545} {\bibfield
  {journal} {\bibinfo  {journal} {Mon. Not. Roy. Astron. Soc.}\ }\textbf
  {\bibinfo {volume} {480}},\ \bibinfo {pages} {302} (\bibinfo {year}
  {2018})},\ \Eprint {http://arxiv.org/abs/1805.00482} {arXiv:1805.00482
  [astro-ph.SR]} \BibitemShut {NoStop}%
\bibitem [{\citenamefont {Finch}\ \emph {et~al.}(2023)\citenamefont {Finch},
  \citenamefont {Bartolucci}, \citenamefont {Chucherko}, \citenamefont
  {Patterson}, \citenamefont {Korol}, \citenamefont {Klein}, \citenamefont
  {Bandopadhyay}, \citenamefont {Middleton}, \citenamefont {Moore},\ and\
  \citenamefont {Vecchio}}]{Finch:2022prg}%
  \BibitemOpen
  \bibfield  {author} {\bibinfo {author} {\bibfnamefont {E.}~\bibnamefont
  {Finch}}, \bibinfo {author} {\bibfnamefont {G.}~\bibnamefont {Bartolucci}},
  \bibinfo {author} {\bibfnamefont {D.}~\bibnamefont {Chucherko}}, \bibinfo
  {author} {\bibfnamefont {B.~G.}\ \bibnamefont {Patterson}}, \bibinfo {author}
  {\bibfnamefont {V.}~\bibnamefont {Korol}}, \bibinfo {author} {\bibfnamefont
  {A.}~\bibnamefont {Klein}}, \bibinfo {author} {\bibfnamefont
  {D.}~\bibnamefont {Bandopadhyay}}, \bibinfo {author} {\bibfnamefont
  {H.}~\bibnamefont {Middleton}}, \bibinfo {author} {\bibfnamefont {C.~J.}\
  \bibnamefont {Moore}}, \ and\ \bibinfo {author} {\bibfnamefont
  {A.}~\bibnamefont {Vecchio}},\ }\href {\doibase 10.1093/mnras/stad1288}
  {\bibfield  {journal} {\bibinfo  {journal} {Mon. Not. Roy. Astron. Soc.}\
  }\textbf {\bibinfo {volume} {522}},\ \bibinfo {pages} {5358} (\bibinfo {year}
  {2023})},\ \Eprint {http://arxiv.org/abs/2210.10812} {arXiv:2210.10812
  [astro-ph.SR]} \BibitemShut {NoStop}%
\bibitem [{\citenamefont {Peters}\ and\ \citenamefont
  {Mathews}(1963)}]{Peters:1963ux}%
  \BibitemOpen
  \bibfield  {author} {\bibinfo {author} {\bibfnamefont {P.~C.}\ \bibnamefont
  {Peters}}\ and\ \bibinfo {author} {\bibfnamefont {J.}~\bibnamefont
  {Mathews}},\ }\href {\doibase 10.1103/PhysRev.131.435} {\bibfield  {journal}
  {\bibinfo  {journal} {Phys. Rev.}\ }\textbf {\bibinfo {volume} {131}},\
  \bibinfo {pages} {435} (\bibinfo {year} {1963})}\BibitemShut {NoStop}%
\bibitem [{\citenamefont {Peters}(1964)}]{Peters:1964zz}%
  \BibitemOpen
  \bibfield  {author} {\bibinfo {author} {\bibfnamefont {P.~C.}\ \bibnamefont
  {Peters}},\ }\href {\doibase 10.1103/PhysRev.136.B1224} {\bibfield  {journal}
  {\bibinfo  {journal} {Phys. Rev.}\ }\textbf {\bibinfo {volume} {136}},\
  \bibinfo {pages} {B1224} (\bibinfo {year} {1964})}\BibitemShut {NoStop}%
\bibitem [{\citenamefont {Cornish}\ and\ \citenamefont
  {Larson}(2003)}]{Cornish:2003vj}%
  \BibitemOpen
  \bibfield  {author} {\bibinfo {author} {\bibfnamefont {N.~J.}\ \bibnamefont
  {Cornish}}\ and\ \bibinfo {author} {\bibfnamefont {S.~L.}\ \bibnamefont
  {Larson}},\ }\href {\doibase 10.1103/PhysRevD.67.103001} {\bibfield
  {journal} {\bibinfo  {journal} {Phys. Rev. D}\ }\textbf {\bibinfo {volume}
  {67}},\ \bibinfo {pages} {103001} (\bibinfo {year} {2003})},\ \Eprint
  {http://arxiv.org/abs/astro-ph/0301548} {arXiv:astro-ph/0301548} \BibitemShut
  {NoStop}%
\bibitem [{\citenamefont {Cornish}\ and\ \citenamefont
  {Littenberg}(2007)}]{Cornish:2007if}%
  \BibitemOpen
  \bibfield  {author} {\bibinfo {author} {\bibfnamefont {N.~J.}\ \bibnamefont
  {Cornish}}\ and\ \bibinfo {author} {\bibfnamefont {T.~B.}\ \bibnamefont
  {Littenberg}},\ }\href {\doibase 10.1103/PhysRevD.76.083006} {\bibfield
  {journal} {\bibinfo  {journal} {Phys. Rev. D}\ }\textbf {\bibinfo {volume}
  {76}},\ \bibinfo {pages} {083006} (\bibinfo {year} {2007})},\ \Eprint
  {http://arxiv.org/abs/0704.1808} {arXiv:0704.1808 [gr-qc]} \BibitemShut
  {NoStop}%
\bibitem [{\citenamefont {Thiele}\ \emph {et~al.}(2023)\citenamefont {Thiele},
  \citenamefont {Breivik}, \citenamefont {Sanderson},\ and\ \citenamefont
  {Luger}}]{Thiele:2021yyb}%
  \BibitemOpen
  \bibfield  {author} {\bibinfo {author} {\bibfnamefont {S.}~\bibnamefont
  {Thiele}}, \bibinfo {author} {\bibfnamefont {K.}~\bibnamefont {Breivik}},
  \bibinfo {author} {\bibfnamefont {R.~E.}\ \bibnamefont {Sanderson}}, \ and\
  \bibinfo {author} {\bibfnamefont {R.}~\bibnamefont {Luger}},\ }\href
  {\doibase 10.3847/1538-4357/aca7be} {\bibfield  {journal} {\bibinfo
  {journal} {Astrophys. J.}\ }\textbf {\bibinfo {volume} {945}},\ \bibinfo
  {pages} {162} (\bibinfo {year} {2023})},\ \Eprint
  {http://arxiv.org/abs/2111.13700} {arXiv:2111.13700 [astro-ph.HE]}
  \BibitemShut {NoStop}%
\bibitem [{\citenamefont {Wagg}\ \emph {et~al.}(2022)\citenamefont {Wagg},
  \citenamefont {Broekgaarden}, \citenamefont {de~Mink}, \citenamefont {van
  Son}, \citenamefont {Frankel},\ and\ \citenamefont {Justham}}]{Wagg:2021cst}%
  \BibitemOpen
  \bibfield  {author} {\bibinfo {author} {\bibfnamefont {T.}~\bibnamefont
  {Wagg}}, \bibinfo {author} {\bibfnamefont {F.~S.}\ \bibnamefont
  {Broekgaarden}}, \bibinfo {author} {\bibfnamefont {S.~E.}\ \bibnamefont
  {de~Mink}}, \bibinfo {author} {\bibfnamefont {L.~A.~C.}\ \bibnamefont {van
  Son}}, \bibinfo {author} {\bibfnamefont {N.}~\bibnamefont {Frankel}}, \ and\
  \bibinfo {author} {\bibfnamefont {S.}~\bibnamefont {Justham}},\ }\href
  {\doibase 10.3847/1538-4357/ac8675} {\bibfield  {journal} {\bibinfo
  {journal} {Astrophys. J.}\ }\textbf {\bibinfo {volume} {937}},\ \bibinfo
  {pages} {118} (\bibinfo {year} {2022})},\ \Eprint
  {http://arxiv.org/abs/2111.13704} {arXiv:2111.13704 [astro-ph.HE]}
  \BibitemShut {NoStop}%
\bibitem [{\citenamefont {Breivik}\ \emph {et~al.}(2018)\citenamefont
  {Breivik}, \citenamefont {Kremer}, \citenamefont {Bueno}, \citenamefont
  {Larson}, \citenamefont {Coughlin},\ and\ \citenamefont
  {Kalogera}}]{Breivik:2017jip}%
  \BibitemOpen
  \bibfield  {author} {\bibinfo {author} {\bibfnamefont {K.}~\bibnamefont
  {Breivik}}, \bibinfo {author} {\bibfnamefont {K.}~\bibnamefont {Kremer}},
  \bibinfo {author} {\bibfnamefont {M.}~\bibnamefont {Bueno}}, \bibinfo
  {author} {\bibfnamefont {S.~L.}\ \bibnamefont {Larson}}, \bibinfo {author}
  {\bibfnamefont {S.}~\bibnamefont {Coughlin}}, \ and\ \bibinfo {author}
  {\bibfnamefont {V.}~\bibnamefont {Kalogera}},\ }\href {\doibase
  10.3847/2041-8213/aaaa23} {\bibfield  {journal} {\bibinfo  {journal}
  {Astrophys. J. Lett.}\ }\textbf {\bibinfo {volume} {854}},\ \bibinfo {pages}
  {L1} (\bibinfo {year} {2018})},\ \Eprint {http://arxiv.org/abs/1710.08370}
  {arXiv:1710.08370 [astro-ph.SR]} \BibitemShut {NoStop}%
\bibitem [{\citenamefont {Tauris}(2018)}]{Tauris:2018kzq}%
  \BibitemOpen
  \bibfield  {author} {\bibinfo {author} {\bibfnamefont {T.~M.}\ \bibnamefont
  {Tauris}},\ }\href {\doibase 10.1103/PhysRevLett.121.131105} {\bibfield
  {journal} {\bibinfo  {journal} {Phys. Rev. Lett.}\ }\textbf {\bibinfo
  {volume} {121}},\ \bibinfo {pages} {131105} (\bibinfo {year} {2018})},\
  \bibinfo {note} {[Erratum: Phys.Rev.Lett. 124, 149902 (2020)]},\ \Eprint
  {http://arxiv.org/abs/1809.03504} {arXiv:1809.03504 [astro-ph.SR]}
  \BibitemShut {NoStop}%
\bibitem [{\citenamefont {Yi}\ \emph {et~al.}(2023)\citenamefont {Yi},
  \citenamefont {Lau}, \citenamefont {Yagi},\ and\ \citenamefont
  {Arras}}]{Yi:2023osk}%
  \BibitemOpen
  \bibfield  {author} {\bibinfo {author} {\bibfnamefont {S.}~\bibnamefont
  {Yi}}, \bibinfo {author} {\bibfnamefont {S.~Y.}\ \bibnamefont {Lau}},
  \bibinfo {author} {\bibfnamefont {K.}~\bibnamefont {Yagi}}, \ and\ \bibinfo
  {author} {\bibfnamefont {P.}~\bibnamefont {Arras}},\ }\href@noop {}
  {\bibfield  {journal} {\bibinfo  {journal} {arXiv}\ } (\bibinfo {year}
  {2023})},\ \Eprint {http://arxiv.org/abs/2310.16172} {arXiv:2310.16172
  [astro-ph.HE]} \BibitemShut {NoStop}%
\bibitem [{\citenamefont {Takahashi}\ and\ \citenamefont
  {Seto}(2002)}]{Takahashi:2002ky}%
  \BibitemOpen
  \bibfield  {author} {\bibinfo {author} {\bibfnamefont {R.}~\bibnamefont
  {Takahashi}}\ and\ \bibinfo {author} {\bibfnamefont {N.}~\bibnamefont
  {Seto}},\ }\href {\doibase 10.1086/341483} {\bibfield  {journal} {\bibinfo
  {journal} {Astrophys. J.}\ }\textbf {\bibinfo {volume} {575}},\ \bibinfo
  {pages} {1030} (\bibinfo {year} {2002})},\ \Eprint
  {http://arxiv.org/abs/astro-ph/0204487} {arXiv:astro-ph/0204487} \BibitemShut
  {NoStop}%
\bibitem [{\citenamefont {Cutler}(1998)}]{Cutler:1997ta}%
  \BibitemOpen
  \bibfield  {author} {\bibinfo {author} {\bibfnamefont {C.}~\bibnamefont
  {Cutler}},\ }\href {\doibase 10.1103/PhysRevD.57.7089} {\bibfield  {journal}
  {\bibinfo  {journal} {Phys. Rev. D}\ }\textbf {\bibinfo {volume} {57}},\
  \bibinfo {pages} {7089} (\bibinfo {year} {1998})},\ \Eprint
  {http://arxiv.org/abs/gr-qc/9703068} {arXiv:gr-qc/9703068} \BibitemShut
  {NoStop}%
\bibitem [{\citenamefont {Berti}\ \emph {et~al.}(2005)\citenamefont {Berti},
  \citenamefont {Buonanno},\ and\ \citenamefont {Will}}]{Berti:2004bd}%
  \BibitemOpen
  \bibfield  {author} {\bibinfo {author} {\bibfnamefont {E.}~\bibnamefont
  {Berti}}, \bibinfo {author} {\bibfnamefont {A.}~\bibnamefont {Buonanno}}, \
  and\ \bibinfo {author} {\bibfnamefont {C.~M.}\ \bibnamefont {Will}},\ }\href
  {\doibase 10.1103/PhysRevD.71.084025} {\bibfield  {journal} {\bibinfo
  {journal} {Phys. Rev. D}\ }\textbf {\bibinfo {volume} {71}},\ \bibinfo
  {pages} {084025} (\bibinfo {year} {2005})},\ \Eprint
  {http://arxiv.org/abs/gr-qc/0411129} {arXiv:gr-qc/0411129} \BibitemShut
  {NoStop}%
\bibitem [{\citenamefont {Robson}\ \emph {et~al.}(2019)\citenamefont {Robson},
  \citenamefont {Cornish},\ and\ \citenamefont {Liu}}]{Robson:2018ifk}%
  \BibitemOpen
  \bibfield  {author} {\bibinfo {author} {\bibfnamefont {T.}~\bibnamefont
  {Robson}}, \bibinfo {author} {\bibfnamefont {N.~J.}\ \bibnamefont {Cornish}},
  \ and\ \bibinfo {author} {\bibfnamefont {C.}~\bibnamefont {Liu}},\ }\href
  {\doibase 10.1088/1361-6382/ab1101} {\bibfield  {journal} {\bibinfo
  {journal} {Class. Quant. Grav.}\ }\textbf {\bibinfo {volume} {36}},\ \bibinfo
  {pages} {105011} (\bibinfo {year} {2019})},\ \Eprint
  {http://arxiv.org/abs/1803.01944} {arXiv:1803.01944 [astro-ph.HE]}
  \BibitemShut {NoStop}%
\bibitem [{\citenamefont {Pierro}\ \emph {et~al.}(2001)\citenamefont {Pierro},
  \citenamefont {Pinto}, \citenamefont {Spallicci}, \citenamefont {Laserra},\
  and\ \citenamefont {Recano}}]{Pierro:2000ej}%
  \BibitemOpen
  \bibfield  {author} {\bibinfo {author} {\bibfnamefont {V.}~\bibnamefont
  {Pierro}}, \bibinfo {author} {\bibfnamefont {I.~M.}\ \bibnamefont {Pinto}},
  \bibinfo {author} {\bibfnamefont {A.~D.}\ \bibnamefont {Spallicci}}, \bibinfo
  {author} {\bibfnamefont {E.}~\bibnamefont {Laserra}}, \ and\ \bibinfo
  {author} {\bibfnamefont {F.}~\bibnamefont {Recano}},\ }\href {\doibase
  10.1046/j.1365-8711.2001.04442.x} {\bibfield  {journal} {\bibinfo  {journal}
  {Mon. Not. Roy. Astron. Soc.}\ }\textbf {\bibinfo {volume} {325}},\ \bibinfo
  {pages} {358} (\bibinfo {year} {2001})},\ \Eprint
  {http://arxiv.org/abs/gr-qc/0005044} {arXiv:gr-qc/0005044} \BibitemShut
  {NoStop}%
\bibitem [{\citenamefont {Isi}(2023)}]{Isi:2022mbx}%
  \BibitemOpen
  \bibfield  {author} {\bibinfo {author} {\bibfnamefont {M.}~\bibnamefont
  {Isi}},\ }\href {\doibase 10.1088/1361-6382/acf28c} {\bibfield  {journal}
  {\bibinfo  {journal} {Class. Quant. Grav.}\ }\textbf {\bibinfo {volume}
  {40}},\ \bibinfo {pages} {203001} (\bibinfo {year} {2023})},\ \Eprint
  {http://arxiv.org/abs/2208.03372} {arXiv:2208.03372 [gr-qc]} \BibitemShut
  {NoStop}%
\bibitem [{lis()}]{lisajous}%
  \BibitemOpen
  \href@noop {} {}\bibinfo {note} {{\texttt{lisajous} code webpage: \\
  \url{https://github.com/cosmoVlad/lisajous} }}\BibitemShut {NoStop}%
\bibitem [{\citenamefont {Sesana}\ \emph {et~al.}(2021)\citenamefont {Sesana}
  \emph {et~al.}}]{Sesana:2019vho}%
  \BibitemOpen
  \bibfield  {author} {\bibinfo {author} {\bibfnamefont {A.}~\bibnamefont
  {Sesana}} \emph {et~al.},\ }\href {\doibase 10.1007/s10686-021-09709-9}
  {\bibfield  {journal} {\bibinfo  {journal} {Exper. Astron.}\ }\textbf
  {\bibinfo {volume} {51}},\ \bibinfo {pages} {1333} (\bibinfo {year}
  {2021})},\ \Eprint {http://arxiv.org/abs/1908.11391} {arXiv:1908.11391
  [astro-ph.IM]} \BibitemShut {NoStop}%
\bibitem [{\citenamefont {Bradbury}\ \emph {et~al.}(2018)\citenamefont
  {Bradbury}, \citenamefont {Frostig}, \citenamefont {Hawkins}, \citenamefont
  {Johnson}, \citenamefont {Leary}, \citenamefont {Maclaurin}, \citenamefont
  {Necula}, \citenamefont {Paszke}, \citenamefont {Vander{P}las}, \citenamefont
  {Wanderman-{M}ilne},\ and\ \citenamefont {Zhang}}]{jax2018github}%
  \BibitemOpen
  \bibfield  {author} {\bibinfo {author} {\bibfnamefont {J.}~\bibnamefont
  {Bradbury}}, \bibinfo {author} {\bibfnamefont {R.}~\bibnamefont {Frostig}},
  \bibinfo {author} {\bibfnamefont {P.}~\bibnamefont {Hawkins}}, \bibinfo
  {author} {\bibfnamefont {M.~J.}\ \bibnamefont {Johnson}}, \bibinfo {author}
  {\bibfnamefont {C.}~\bibnamefont {Leary}}, \bibinfo {author} {\bibfnamefont
  {D.}~\bibnamefont {Maclaurin}}, \bibinfo {author} {\bibfnamefont
  {G.}~\bibnamefont {Necula}}, \bibinfo {author} {\bibfnamefont
  {A.}~\bibnamefont {Paszke}}, \bibinfo {author} {\bibfnamefont
  {J.}~\bibnamefont {Vander{P}las}}, \bibinfo {author} {\bibfnamefont
  {S.}~\bibnamefont {Wanderman-{M}ilne}}, \ and\ \bibinfo {author}
  {\bibfnamefont {Q.}~\bibnamefont {Zhang}},\ }\href
  {http://github.com/google/jax} {\enquote {\bibinfo {title} {{JAX}: composable
  transformations of {P}ython+{N}um{P}y programs},}\ } (\bibinfo {year}
  {2018})\BibitemShut {NoStop}%
\bibitem [{\citenamefont {{Perez}}\ and\ \citenamefont
  {{Granger}}(2007)}]{2007CSE.....9c..21P}%
  \BibitemOpen
  \bibfield  {author} {\bibinfo {author} {\bibfnamefont {F.}~\bibnamefont
  {{Perez}}}\ and\ \bibinfo {author} {\bibfnamefont {B.~E.}\ \bibnamefont
  {{Granger}}},\ }\href {\doibase 10.1109/MCSE.2007.53} {\bibfield  {journal}
  {\bibinfo  {journal} {Computing in Science and Engineering}\ }\textbf
  {\bibinfo {volume} {9}},\ \bibinfo {pages} {21} (\bibinfo {year}
  {2007})}\BibitemShut {NoStop}%
\bibitem [{\citenamefont {{Virtanen}}\ \emph {et~al.}(2020)\citenamefont
  {{Virtanen}}, \citenamefont {{Gommers}}, \citenamefont {{Oliphant}},
  \citenamefont {{Haberland}}, \citenamefont {{Reddy}}, \citenamefont
  {{Cournapeau}}, \citenamefont {{Burovski}}, \citenamefont {{Peterson}},
  \citenamefont {{Weckesser}}, \citenamefont {{Bright}}, \citenamefont {{van
  der Walt}}, \citenamefont {{Brett}}, \citenamefont {{Wilson}}, \citenamefont
  {{Millman}}, \citenamefont {{Mayorov}}, \citenamefont {{Nelson}},
  \citenamefont {{Jones}}, \citenamefont {{Kern}}, \citenamefont {{Larson}},
  \citenamefont {{Carey}}, \citenamefont {{Polat}}, \citenamefont {{Feng}},
  \citenamefont {{Moore}}, \citenamefont {{VanderPlas}}, \citenamefont
  {{Laxalde}}, \citenamefont {{Perktold}}, \citenamefont {{Cimrman}},
  \citenamefont {{Henriksen}}, \citenamefont {{Quintero}}, \citenamefont
  {{Harris}}, \citenamefont {{Archibald}}, \citenamefont {{Ribeiro}},
  \citenamefont {{Pedregosa}}, \citenamefont {{van Mulbregt}},\ and\
  \citenamefont {{SciPy 1. 0 Contributors}}}]{2020NatMe..17..261V}%
  \BibitemOpen
  \bibfield  {author} {\bibinfo {author} {\bibfnamefont {P.}~\bibnamefont
  {{Virtanen}}}, \bibinfo {author} {\bibfnamefont {R.}~\bibnamefont
  {{Gommers}}}, \bibinfo {author} {\bibfnamefont {T.~E.}\ \bibnamefont
  {{Oliphant}}}, \bibinfo {author} {\bibfnamefont {M.}~\bibnamefont
  {{Haberland}}}, \bibinfo {author} {\bibfnamefont {T.}~\bibnamefont
  {{Reddy}}}, \bibinfo {author} {\bibfnamefont {D.}~\bibnamefont
  {{Cournapeau}}}, \bibinfo {author} {\bibfnamefont {E.}~\bibnamefont
  {{Burovski}}}, \bibinfo {author} {\bibfnamefont {P.}~\bibnamefont
  {{Peterson}}}, \bibinfo {author} {\bibfnamefont {W.}~\bibnamefont
  {{Weckesser}}}, \bibinfo {author} {\bibfnamefont {J.}~\bibnamefont
  {{Bright}}}, \bibinfo {author} {\bibfnamefont {S.~J.}\ \bibnamefont {{van der
  Walt}}}, \bibinfo {author} {\bibfnamefont {M.}~\bibnamefont {{Brett}}},
  \bibinfo {author} {\bibfnamefont {J.}~\bibnamefont {{Wilson}}}, \bibinfo
  {author} {\bibfnamefont {K.~J.}\ \bibnamefont {{Millman}}}, \bibinfo {author}
  {\bibfnamefont {N.}~\bibnamefont {{Mayorov}}}, \bibinfo {author}
  {\bibfnamefont {A.~R.~J.}\ \bibnamefont {{Nelson}}}, \bibinfo {author}
  {\bibfnamefont {E.}~\bibnamefont {{Jones}}}, \bibinfo {author} {\bibfnamefont
  {R.}~\bibnamefont {{Kern}}}, \bibinfo {author} {\bibfnamefont
  {E.}~\bibnamefont {{Larson}}}, \bibinfo {author} {\bibfnamefont {C.~J.}\
  \bibnamefont {{Carey}}}, \bibinfo {author} {\bibfnamefont
  {{\.I}.}~\bibnamefont {{Polat}}}, \bibinfo {author} {\bibfnamefont
  {Y.}~\bibnamefont {{Feng}}}, \bibinfo {author} {\bibfnamefont {E.~W.}\
  \bibnamefont {{Moore}}}, \bibinfo {author} {\bibfnamefont {J.}~\bibnamefont
  {{VanderPlas}}}, \bibinfo {author} {\bibfnamefont {D.}~\bibnamefont
  {{Laxalde}}}, \bibinfo {author} {\bibfnamefont {J.}~\bibnamefont
  {{Perktold}}}, \bibinfo {author} {\bibfnamefont {R.}~\bibnamefont
  {{Cimrman}}}, \bibinfo {author} {\bibfnamefont {I.}~\bibnamefont
  {{Henriksen}}}, \bibinfo {author} {\bibfnamefont {E.~A.}\ \bibnamefont
  {{Quintero}}}, \bibinfo {author} {\bibfnamefont {C.~R.}\ \bibnamefont
  {{Harris}}}, \bibinfo {author} {\bibfnamefont {A.~M.}\ \bibnamefont
  {{Archibald}}}, \bibinfo {author} {\bibfnamefont {A.~H.}\ \bibnamefont
  {{Ribeiro}}}, \bibinfo {author} {\bibfnamefont {F.}~\bibnamefont
  {{Pedregosa}}}, \bibinfo {author} {\bibfnamefont {P.}~\bibnamefont {{van
  Mulbregt}}}, \ and\ \bibinfo {author} {\bibnamefont {{SciPy 1. 0
  Contributors}}},\ }\href {\doibase 10.1038/s41592-019-0686-2} {\bibfield
  {journal} {\bibinfo  {journal} {Nature Methods}\ }\textbf {\bibinfo {volume}
  {17}},\ \bibinfo {pages} {261} (\bibinfo {year} {2020})},\ \Eprint
  {http://arxiv.org/abs/1907.10121} {arXiv:1907.10121 [cs.MS]} \BibitemShut
  {NoStop}%
\bibitem [{\citenamefont {{Hunter}}(2007)}]{2007CSE.....9...90H}%
  \BibitemOpen
  \bibfield  {author} {\bibinfo {author} {\bibfnamefont {J.~D.}\ \bibnamefont
  {{Hunter}}},\ }\href {\doibase 10.1109/MCSE.2007.55} {\bibfield  {journal}
  {\bibinfo  {journal} {Computing in Science and Engineering}\ }\textbf
  {\bibinfo {volume} {9}},\ \bibinfo {pages} {90} (\bibinfo {year}
  {2007})}\BibitemShut {NoStop}%
\bibitem [{\citenamefont {{van der Walt}}\ \emph {et~al.}(2011)\citenamefont
  {{van der Walt}}, \citenamefont {{Colbert}},\ and\ \citenamefont
  {{Varoquaux}}}]{2011CSE....13b..22V}%
  \BibitemOpen
  \bibfield  {author} {\bibinfo {author} {\bibfnamefont {S.}~\bibnamefont {{van
  der Walt}}}, \bibinfo {author} {\bibfnamefont {S.~C.}\ \bibnamefont
  {{Colbert}}}, \ and\ \bibinfo {author} {\bibfnamefont {G.}~\bibnamefont
  {{Varoquaux}}},\ }\href {\doibase 10.1109/MCSE.2011.37} {\bibfield  {journal}
  {\bibinfo  {journal} {Computing in Science and Engineering}\ }\textbf
  {\bibinfo {volume} {13}},\ \bibinfo {pages} {22} (\bibinfo {year} {2011})},\
  \Eprint {http://arxiv.org/abs/1102.1523} {arXiv:1102.1523 [cs.MS]}
  \BibitemShut {NoStop}%
\bibitem [{\citenamefont {Meurer}\ \emph {et~al.}(2017)\citenamefont {Meurer}
  \emph {et~al.}}]{Meurer:2017yhf}%
  \BibitemOpen
  \bibfield  {author} {\bibinfo {author} {\bibfnamefont {A.}~\bibnamefont
  {Meurer}} \emph {et~al.},\ }\href {\doibase 10.7717/peerj-cs.103} {\bibfield
  {journal} {\bibinfo  {journal} {PeerJ Comput. Sci.}\ }\textbf {\bibinfo
  {volume} {3}},\ \bibinfo {pages} {e103} (\bibinfo {year} {2017})}\BibitemShut
  {NoStop}%
\bibitem [{\citenamefont {Johansson}\ \emph {et~al.}(2013)\citenamefont
  {Johansson} \emph {et~al.}}]{mpmath}%
  \BibitemOpen
  \bibfield  {author} {\bibinfo {author} {\bibfnamefont {F.}~\bibnamefont
  {Johansson}} \emph {et~al.},\ }\href@noop {} {\enquote {\bibinfo {title}
  {mpmath: a {P}ython library for arbitrary-precision floating-point arithmetic
  (version 1.1.0)},}\ } (\bibinfo {year} {2013}),\ \bibinfo {note} {{\tt
  http://mpmath.org/}}\BibitemShut {NoStop}%
\bibitem [{\citenamefont {{Gerosa}}\ and\ \citenamefont
  {{Vallisneri}}(2017)}]{2017JOSS....2..222G}%
  \BibitemOpen
  \bibfield  {author} {\bibinfo {author} {\bibfnamefont {D.}~\bibnamefont
  {{Gerosa}}}\ and\ \bibinfo {author} {\bibfnamefont {M.}~\bibnamefont
  {{Vallisneri}}},\ }\href {\doibase 10.21105/joss.00222} {\bibfield  {journal}
  {\bibinfo  {journal} {The Journal of Open Source Software}\ }\textbf
  {\bibinfo {volume} {2}},\ \bibinfo {pages} {222} (\bibinfo {year}
  {2017})}\BibitemShut {NoStop}%
\end{thebibliography}%

\end{document}